\documentclass[letterpaper,conference]{IEEEtran}

\usepackage[utf8]{inputenc}
\usepackage[T1]{fontenc}
\usepackage{verbatim} 
\usepackage{comment}

\usepackage{cite}

\usepackage{dblfloatfix}
\ifCLASSINFOpdf
  \usepackage[pdftex]{graphicx}
  \graphicspath{{./}}
  \DeclareGraphicsExtensions{pdf,.eps}
\else
  \usepackage[dvips]{graphicx}
  \graphicspath{{./figures/}}
  \DeclareGraphicsExtensions{.eps}
\fi

\usepackage[cmex10]{amsmath}
\usepackage{eucal}          

\usepackage{algorithmic}

\usepackage{array}
\usepackage{mdwmath}
\usepackage{mdwtab}

\usepackage[tight,footnotesize]{subfigure}

\usepackage{url}

\usepackage{fancyhdr}
\usepackage{amsfonts}
\usepackage{amsthm}
\usepackage{tabularx}
\usepackage[printonlyused]{acronym}
\usepackage{framed}
\usepackage{algorithm}
\usepackage{flushend}

\usepackage{nccmath}
\usepackage[draft,footnote,nomargin]{fixme}

\newcommand{\tn}{\textnormal}

\newcolumntype{v}[1]{>{\raggedright\hspace{0pt}}p{#1}}
\renewcommand{\arraystretch}{1.2}

\newcommand{\fref}[1]{Fig.~\ref{#1}}
\newcommand{\sref}[1]{Section~\ref{#1}}
\newcommand{\tref}[1]{Table~\ref{#1}}
\newcommand{\aref}[1]{Algorithm~\ref{#1}}
\newcommand{\eref}[1]{(\ref{#1})}

\begin{document}

\acrodef{ADB}{Android Debug Bridge}
\acrodef{AP}{Access Point}
\acrodef{HLS}{HTTP Live Streaming}
\acrodef{PHY}{Physical layer}
\acrodef{ICT}{Information and Communication Technology}
\acrodef{PGS}{Predictive Green Streaming}
\acrodef{AVS}{Adaptive video streaming}
\acrodef{BS}{Base Station}
\acrodef{ES}{Equal Share}
\acrodef{RP}{Rate-Proportional}
\acrodef{RA}{Rate Allocation}
\acrodef{LHS}{left hand side}
\acrodef{SON}{Self Organizing Networks}
\acrodef{RHS}{right hand side}
\acrodef{PD}{Progressive Download}
\acrodef{VD}{Video Degradation}
\acrodef{QD}{Quality Degradation}
\acrodef{RR}{Round-robin}
\acrodef{SNR}{Signal-to-Noise Ratio}
\acrodef{CQI}{Channel Quality Information}
\acrodef{CoMP}{Coordinated Multi-Point}
\acrodef{LP}{Linear Program}
\acrodef{DASH}{Dynamic Adaptive Streaming over HTTP}
\acrodef{MILP}{Mixed Integer Linear Program}
\acrodef{LIRA}{Long-term Inter-cell Resource Allocation}
\acrodef{PASS}{Predictive Adaptive Stored Streaming}
\acrodef{PAS}{Predictive Adaptive Streaming}
\acrodef{LLS}{Long-term Lookback Scheduling}
\acrodef{LL}{Long-term Lookback}
\acrodef{LLPF}{Long-term Lookback Proportional Fair}
\acrodef{LL-EXP}{Long-term Lookback Exponential}
\acrodef{LL-PF}{Long-term Lookback Proportional Fair}
\acrodef{EXP-F}{Exponential with Freezing}
\acrodef{EXP}{Exponential}
\acrodef{AS}{Adaptive Streaming}
\acrodef{QoE}{Quality of Experience}
\acrodef{3GPP}{3rd Generation Partnership Project}
\acrodef{DRX}{Discontinuous Reception}
\acrodef{QoS}{Quality of Service}
\acrodef{MA}{Moving Average}
\acrodef{MR}{Maximum Rate}
\acrodef{RWP}{Random Way Point}
\acrodef{PF}{Proportional Fair}
\acrodef{RA}{Resource Allocation}
\acrodef{PRA}{Predictive Resource Allocation}
\acrodef{PFS}{Proportional Fair Scheduling}
\acrodef{MPFS}{Multi-cell Proportional Fair Scheduling}
\acrodef{PRB}{Physical Resource Block}
\acrodef{IPF}{Inter-cell Proportional Fair}
\acrodef{MPF}{Multi-cell Proportional Fair}
\acrodef{MSE}{Mean Square Error}
\acrodef{SPF}{Single-cell Proportional Fair}
\acrodef{SPFS}{Single-cell Proportional Fair Scheduling}
\acrodef{QoS}{Quality of Service}
\acrodef{TCP}{Transmission Control Protocol}
\acrodef{TTI}{Time Transmission Interval}
\acrodef{UE}{User Equipment}
\acrodef{TDMA}{Time Division Multiple Access}
\acrodef{MAC}{Medium Access Control}
    \acrodef{LTE}{Long Term Evolution}
    \acrodef{HARQ}{Hybrid Automatic Repeat Request}
    \acrodef{STC}{Space-Time Coding}
    \acrodef{QAM}{Quadrature Amplitude Modulation}

\bibliographystyle{ieeetr}

\IEEEoverridecommandlockouts	

\title{Energy-Efficient Adaptive Video Transmission: Exploiting Rate Predictions in Wireless Networks}

\author{Hatem~Abou-zeid,~\IEEEmembership{Student Member,~IEEE,}
        Hossam~S.~Hassanein,~\IEEEmembership{Senior Member,~IEEE,}\\
        and~Stefan~Valentin,~\IEEEmembership{Member,~IEEE}
\thanks{H. Abou-zeid is with the Department of Electrical and Computer Engineering, Queen's University, Kingston,  Canada, E-mail: h.abouzeid@queensu.ca.}
\thanks{H.~S.~Hassanein is with the School of Computing, Queen's University, Kingston,
Canada, E-mail: hossam@cs.queensu.ca.}
\thanks{S. Valentin is with Bell Labs,  Alcatel-Lucent, Germany, E-mail: stefan.valentin@alcatel-lucent.com.}
\thanks{Copyright (c) 2013 IEEE. Personal use of this material is permitted. However, permission to use this material for any other purposes must be obtained from the IEEE by sending a request to pubs-permissions@ieee.org.}  }

\maketitle

\begin{abstract}
The unprecedented growth of mobile video traffic is adding significant pressure to the energy drain at both the network and the end user. Energy efficient video transmission techniques are thus imperative to cope with the challenge of satisfying user demand at sustainable costs. In this paper, we investigate how predicted user rates can be exploited for energy-efficient video streaming with the popular HTTP-based \ac{AS} protocols (e.g. DASH). 
To this end, we develop an energy-efficient \ac{PGS} optimization framework that leverages predictions of wireless data rates to achieve the following objectives 1) minimize the required transmission airtime without causing streaming interruptions, 2) minimize total downlink \ac{BS} power consumption for cases where BSs can be switched off in deep sleep, and 3) enable a trade-off between \ac{AS} quality and energy consumption.
Our framework is first formulated as a \ac{MILP} where decisions on multi-user rate allocation, video segment quality, and BS transmit power are \textit{jointly} optimized. Then, to provide an online solution, we present a polynomial-time heuristic algorithm that decouples the \ac{PGS} problem into multiple stages. We provide
a performance analysis of the proposed methods by simulations, and numerical results demonstrate that the \ac{PGS} framework yields significant energy savings.

\end{abstract}
\medskip

\section{Introduction}
\label{sec:intro}

The increasing mobile data traffic and dense deployment of \acfp{BS} have made energy efficiency in networks imperative. This traffic growth is not only adding more pressure to the network and user device energy drain, but also increases network operational expenditures (OPEX) and  negatively impacts the environment \cite{greenSurvey:Hasan2011}. Consequently, research and standardization efforts are focusing on devising green mechanisms to save energy across the network. Among the wireless network elements, \acp{BS} account for more than 50\,\% of the network energy consumption \cite{greenAccess:Chen}. Reducing BS downlink transmit power by efficient rate allocation will thus result in monetary savings for the operator, and also reduce $\tn{CO}_2$ emissions. Furthermore, energy-efficient rate allocation can improve the spectral efficiency and provide additional resources at high demand. Therefore, devising green radio \emph{access} strategies is vital for overall network performance. 

Meanwhile, video streaming is experiencing unprecedented growth with forecasts indicating that it will soon account for 66\,\% of the total mobile traffic \cite{VNI}. This is driven by the increasing capabilities of mobile devices and by content aggregation sites such as Netflix and YouTube.
In particular, \acf{AS} is gaining popularity due to its ability to seamlessly adapt streaming quality to the current wireless data rate. In HTTP-based implementations, such as \ac{HLS} \cite{hls}, or \ac{DASH} \cite{iso_dash}, the video content is divided into a sequence of small file segments, each containing a short interval of playback time. Each segment is made available at multiple bitrates, and depending on the network conditions, the suitable segment quality is selected for transmission \cite{Begen:DASH}. This reduces video freezing and is particularly suited for mobile video streaming where users experience channel gain fluctuations. While \ac{AS} improves user \ac{QoS}, energy consumption still remains a fundamental challenge. 

In this paper, we investigate how predictions of user rates can be exploited for energy efficient transmission of stored videos that can be strategically buffered at the user devices. The predictability of a wireless channel is generally possible due to the correlation between location and channel capacity \cite{WirelessPred:Malmirchengini12},\cite{Yao:BWPredictability}. Therefore, if a user's future location is known, the upcoming data rates can be anticipated from radio and coverage maps stored at the network, which can also be updated in real-time from \ac{UE} measurements \cite{driveTests:Johansson2012}.
While mobility predictions are particularly plausible for users commuting in public transportation, trains, or vehicles on highways \cite{chen13:localization_ml}, studies on human mobility patterns reveal a high degree of temporal and spatial regularity \cite{mobilitypatterns}, suggesting a potential 93\% predictability \cite{mobilitypredictability}.
A key motivation for incorporating such predictions is the plethora of navigation and context information available in today's smart phones.

Being aware of a user's upcoming rate allows the network to plan spectrally efficient rate allocations in advance without violating user streaming demands. For instance, if a user is moving towards the cell edge or a tunnel, the network can increase the allocated wireless resources allowing the user to buffer more video segments. \emph{Pre-buffering} this additional data then provides smooth video streaming since the user can consume the buffer while being in poor radio coverage. Additionally, by not serving users in such conditions, the network-wide spectral efficiency increases since valuable channel resources can be provided to other users instead. If, on the other hand, the user is approaching the \ac{BS}, transmission can be delayed, provided sufficient video segments have previously been buffered. This allows the BS to save energy by `sleeping' as the user approaches, and then `waking up' for a short period, during which a high data transmission is possible. 

We summarize the main contributions of this paper in the following.
\begin{itemize}
\item We propose a \acf{PGS} optimization framework that exploits rate predictions over a time horizon to i) minimize the required transmission airtime subject to a target average quality level, without causing streaming interruptions, ii) minimize total downlink \ac{BS} power consumption for cases where BSs can be switched off in deep sleep, and iii) enable a trade-off between video quality and energy consumption.
\item The \ac{PGS} problem is formulated as an \ac{MILP} that jointly determines multi-user \ac{RA}, video segment quality levels, and BS on/off status. The proposed formulation captures i) the joint relationship between \emph{cumulative} user \ac{RA} and \emph{long-term} segment quality planning, and ii) the load dependent \ac{BS} power consumption, with a minimum off duration for deep sleep modes.
\item For online implementation, we present a polynomial time algorithm that solves the \ac{PGS} problem. Results demonstrate that the proposed algorithm stays close to the MILP benchmark in energy consumption, while exhibiting higher \ac{QoS} \emph{robustness} to rate prediction errors compared to the \ac{MILP}.
\end{itemize}

We compare the performance and robustness of the \ac{PGS} approaches through extensive simulation under realistic assumptions on cellular networks and vehicular mobility. 
We observe up to 85\% energy reduction, while achieving comparable \ac{QoS}, with respect to baseline solutions.
Our results demonstrate that \ac{PGS} is a promising energy saving framework for future cellular networks.

The rest of this paper is organized as follows. We review related work in \sref{sec:RelatedWork}, and introduce the system model in \sref{sec:SystemModel}. The \ac{MILP} formulation of the \ac{PGS} framework is developed in \sref{sec:PGS-MILP}, while \sref{sec:PGS-Heuristic} presents the proposed \ac{PGS} algorithm. In \sref{sec:Sim}, we present simulation results to study the power consumption and video quality performance of \ac{PGS}, and its robustness to prediction errors. Finally, conclusions are given in \sref{sec:Conc}.

\section{Background and Related Work}
\label{sec:RelatedWork}
This paper addresses the problem of energy efficient downlink transmission for adaptive video streaming, in a multi-cell network. In this section, we first provide a background on \emph{traffic-aware} energy efficient radio access, and then discuss related works that exploit rate predictions in detail. 

\subsection{Traffic-Aware Energy Efficient Radio Access}
As networks are over dimensioned to serve peak user demands, radio access energy can be reduced in a number of ways at times of lower demand. 
Such mechanisms include: 1) \textit{time}-domain duty-cycling \cite{greenAccess:Chen},~\cite {Deng:2012} that utilizes only a fraction of the transmission slots and puts the \ac{BS} at low energy operating modes during times of inactivity, and 2) \textit{frequency}-domain duty cycling where only a fraction of the bandwidth (or physical resource blocks) is used for transmission \cite{C.Desset:DTXPowerModel}. 
Additionally, when demand is low for prolonged periods, BSs can be powered down to deep sleep modes that consume negligible power \cite{greenAccessChallenges:Correia2010},~\cite{sleepSmall:Ashraf2011}. 
Information on the temporal and spatial user traffic demand, can assist networks to make better adaptations that reduce energy consumption.
Such traffic-awareness is incorporated in \cite{greenCoop2:Ismail2011} where an optimal on-off switching framework is developed to maximize energy savings under service constraints. More recently, in \cite{Han12:EE} and \cite{greenCoop3:Ansari2013}, multi-cell cooperation is proposed to configure the network layout by powering down select \acp{BS} depending on network traffic. While the preceding works focus on traffic-aware energy efficiency in general, they do not investigate using predictions to reduce BS utilization and energy. 

\subsection{Exploiting Mobility-Based Rate Predictions}

The potential of \emph{mobility-based} rate predictions is receiving increasing interest in recent literature. Ali \emph{et al.} \cite{MAOS} show how the system throughput can be increased with such predictions, while \cite{abouzeid13:PRA, abouzeid14:PPF_GCC, PredPFINFOCOM:Robert} discuss improvements in fairness as well under more realistic evaluation scenarios. Margolies  \emph{et al.} \cite{PredPFINFOCOM:Robert} also use extensive channel measurements from a 3G network to show that a user's channel state is highly reproducible.

Leveraging rate predictions for wireless video streaming has been discussed in \cite{Yao:QoSBWmaps,PredStreaming:Curcio2010, Riiser:Streaming2012}. 
Yao \emph{et al.} \cite{Yao:QoSBWmaps} develop a rate adaptation algorithm that proactively switches to the predicted transmission rates. This improves TCP rate control and throughput by faster convergence to the available capacity, but does not pre-buffer segments or adapt quality based on predictions. This is addressed in \cite{PredStreaming:Curcio2010},\cite{Riiser:Streaming2012} where users heading to poor conditions request additional segments in advance. A prototype is presented in \cite{Riiser:Streaming2012} that logs receiver bandwidth-location information to perform long-term bitrate planning and prevent streaming disruptions.
While these works assume that the user trajectory is known, a related \emph{geo-predictive} quality adaptation mechanism for \ac{DASH} has also recently been presented in \cite{PreVideo:GtubeHao2014}, where the user path is also predicted. 
In our own work \cite{abouzeid14:LRA_CCNC}, we also study how video interruptions can be minimized by exploiting rate predictions. However, the focus of this work is on optimizing \emph{multi-user} resource allocation and not on adapting video quality, as in \cite{Yao:QoSBWmaps,PredStreaming:Curcio2010, Riiser:Streaming2012,PreVideo:GtubeHao2014}, where each client controls its bitrate plan independently. This 
\emph{in-network} resource allocation facilitates obtaining network wide objectives and efficiently trading-off video quality among multiple users. Several recent resource management approaches for video streaming have also been proposed in \cite{schedulingDASH:chen2013, DASH:BellLabs2013, Nova:INFOCOM}, but predictions are not considered therein. 
The aforementioned works focus on enhancing user experience but do not address energy efficiency. 


The work in references \cite{Z.Lu:PredVideoINFOCOM, abouzeid13:PreGWA, abouzeid14:EEVideoWCNC}
is closest to this paper, where the primary objective is to exploit predictions for \emph{energy efficiency}. 
Rate predictions are used to minimize system utilization and avoid streaming delays of constant bit rate videos in \cite{Z.Lu:PredVideoINFOCOM}. The authors present a detailed buffer model and formulate the multi-user single cell case, as a non-convex problem. Then, optimal algorithms for the single user case are developed and significant reductions in BS resource utilization are observed. 
In \cite{abouzeid13:PreGWA}, we also discuss the potential energy savings that can be achieved by a \textit{mobility-aware} wireless access framework. An architecture is presented with the composite functional elements, and their interaction is discussed. 
The work in \cite{abouzeid14:EEVideoWCNC}  considers the problem of trading-off video degradations with BS power consumption, and presents predictive algorithms to solve the problem. 
This paper differs from these works in several aspects,
1) we now consider the delivery of adaptive video streams and, thus, model and solve the \emph{joint} rate allocation-quality planning problem over a time horizon,
2) in addition to saving power by minimizing utilization, we also incorporate deep sleep modes where BSs can be switched off, and 
3) we formulate a detailed multi-user, multi-cell optimization framework for energy efficient \ac{AS}, and present an efficient heuristic algorithm to solve the problem.

\begin{figure*}[!b]
\noindent
\centering
	\subfigure[The defined time durations and slot indices.]{
	\includegraphics[width=65mm]{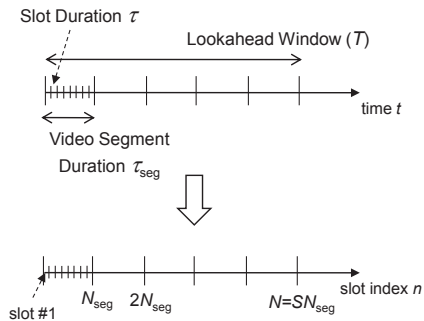}\label{fig:TimeIntervals}}
	\qquad \qquad
	\subfigure[Sample user rate allocation during $T$.]{
	\includegraphics[width=60mm]{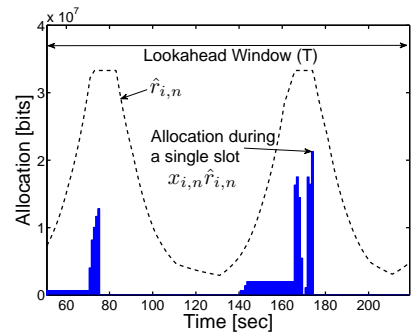}\label{fig:AllocationExample}}
	\caption{System models and notation.}
	\label{fig:ResourceModels}
\end{figure*}
\section{System Model and Preliminaries}
\label{sec:SystemModel}
In this section, we present the system model and assumptions. We use the following notational conventions: $\mathcal{X}$ denotes a set and its cardinality $|\mathcal{X}|$ is denoted by $X$. We use bold letters to denote matrices, e.g. $\mathbf{x}=(x_{a,b}: a \in \mathbb{Z}_+, b \in \mathbb{Z}_+)$.
$\lfloor\cdot\rfloor$ and $\lceil\cdot\rceil$ are the floor and ceiling functions respectively. 
Frequently used notation is summarized in \tref{tab:notation}.
\renewcommand{\arraystretch}{.9}
\begin{table*}[!t]
\caption{Summary of Important Symbols}
\label{tab:notation}
\centering
\begin{tabularx}{1.5\columnwidth}{lX}
\hline
Symbol            & Description \\ \hline
$b_{k,n}$		  & Binary decision variable for on/off status of \ac{BS} $k$ at slot $n$ \\ 
$\tn{BS}^\tn{air}_{k,n}$		  & Available airtime of \ac{BS} $k$ at slot $n$ \\
$D_{i,s}$	& Cumulative number of bits required by user $i$ to stream the first $s$ segments \,[bits]\\ 
$i$		  & User index, $i=\{1,2,\cdots, M\}$\\
$k$		  & BS index, $k=\{1,2,\cdots, K\}$\\ 
$\mathcal{K}$		  & Set of \ac{BS}s in the network \\ 
$\mathcal{M}$		  & Set of users in the network \\ 
$n$		  & Time slot index, $n=\{1,2,\cdots, N\}$\\ 
$N$	& Number of slots in the lookahead window  \\ 
$N_\tn{seg}$	& Number of slots in a video segment \\
$\mathcal{N}$		  & Set of time slots in the lookahead window\\
$\mathcal{N}^s$		  & Set of time slots belonging to segment $s$\\
 $p_{k,n}$		  & Transmit power of \ac{BS} $k$ at slot $n$ \\ 
$q_{i,s,l}$		  & Binary variable for quality level $l$ of segment $s$ for user $i$\\ 
$q_\tn{max}$		  & Maximum quality level \\ 
$\hat r_{i,n}$		  & Link rate of user $i$ at slot $n$\,[bits] \\ 
$R_{i,n}$	& Cumulative rate allocated to user $i$ by slot $n$\,[bits]\\ 
$s$		  & Segment index, $s=\{1,2,\cdots, S\}$\\ 
$S$	& Number of segments in the lookahead window \\ 
$T$	& Duration of the lookahead window [s] \\ 
$\tau$	& Duration of a time slot [s] \\ 
$\tau_\tn{seg}$	& Duration of a video segment [s] \\ 
$\mathcal{U}_{k,n}$ & Set containing the indices of users associated with BS $k$ at slot $n$\\
$x_{i,n}$		  & Fraction of airtime assigned to user $i$ at slot $n$ \\

\hline
\end{tabularx}
\end{table*}
\subsection{Network Overview}
Consider a network with a \ac{BS} set $\mathcal{K}$ and an active user set $\mathcal{M}$. Users request stored video content that is transmitted using adaptive bitrate streaming over HTTP.
Time is divided in slots of equal duration $\tau$, during which the wireless channel can be shared arbitrarily among multiple users. 
We assume that the wireless link is the bottleneck, and therefore the requested video content is always available at the BS for transmission.

\subsection{Link Model and Resource Sharing}
We assume a block fading model where the achievable data rate is assumed to be constant during each time slot.
As we are interested in slow-fading variations, a typical value of such a coherence time $\tau$ is $1$\,s for vehicle speeds up to $20\,$m/s, during which average wireless capacity is not significantly affected. The achievable data rate depends on the path loss model
${\tn{PL}(d)=128.1+37.6\log_{10} d}$, where the user-BS distance $d$ is in km \cite{3GPPSpecRF}. The feasible link rate is computed using Shannon's equation with SNR clipping at 20\,dB to account for practical modulation orders. Therefore, a user $i$ at slot $n$, will have a feasible data rate of 
\begin{equation}
\label{eq:PredRate}
r_{i,n}= \tau B\log_2 (1 + \frac{P_{\tn{rx}_{i,n}}}{N_o B})\quad [\tn{bits}], 
\end{equation}
where $P_\tn{rx}$, $N_o$ and $B$ are the received power, noise power spectral density, and the transmission bandwidth respectively\footnote{This path-loss dependent link model is an abstraction of a radio environment map which will in practice be available at the service provider.}.
Note that the slot user rate $r_{i,n}$ gives the number of bits that can be transmitted during a time slot, i.e. the transmission rate normalized with slot duration $\tau$.

User link capacities are assumed to be known for the upcoming $T$ seconds, which we call the \textit{lookahead window}. There are 
$N=T/\tau$ time slots within the lookahead window as illustrated in \fref{fig:TimeIntervals}, which we denote by the set $\mathcal{N}=\{1,2,\cdots,N\}$. The future link capacities are determined by computing $P_{\tn{rx}_{i,n}}$ based on the knowledge of the future user locations, and then substituting in \eref{eq:PredRate}. This will generate a matrix of future link rates as defined by $\mathbf{\hat r}=(\hat r_{i,n}: {i\in\mathcal{M}},{n\in \mathcal{N}}).$
\fref{fig:AllocationExample} illustrates an example of $\hat{r}_{i,n},\forall n$ for a user traversing two BSs along a highway. We first assume that knowledge of $\mathbf{\hat r}$ is error free to provide the bounds of the potential gains, and then assess the impact of prediction errors on the gains.

\fref{fig:TimeIntervals} also illustrates the video segment duration $\tau_\tn{seg}$, which is a multiple of $\tau$. The lookahead window $T$ is also selected to be divisible by $\tau_\tn{seg}$ as shown in the figure. In terms of time slots, $N_\tn{seg}=\tau_\tn{seg}/\tau$ denotes the number of slots that make up one video segment, and $S=N/N_\tn{seg}$ denotes the number of video segments during $T$.

BS airtime is shared among the active users during each slot $n$. We define the rate allocation matrix $\mathbf{x}=(x_{i,n} \in [0,1]: {i\in\mathcal{M}},{n\in \mathcal{N}})$ which gives the fraction of time during each slot $n$ that the BS bandwidth is assigned to user $i$. The rate received by each user at each slot is the element-wise product $\mathbf{x}\odot\mathbf{\hat{r}}$. Therefore, $\mathbf{x}$ controls both the \textit{per-slot} and total \textit{long-term} rates users receive over the $N$ slots. A sample allocation $x_{i,n},\forall n$ for a user $i$ is illustrated in \fref{fig:AllocationExample}, where the bars indicate the proportion of $\hat{r}_{i,n}$ allocated to that user. Note that, since a user can traverse multiple cells during $N$, BS cooperation is needed to make the allocation plan. This is assumed to be possible via an inter-BS interface such as the X2-interface in \ac{3GPP} compliant networks. User-BS association is based on the strongest received signal. 
We can define the set $\mathcal{U}_{k,n},k \in \mathcal{K},{n\in \mathcal{N}}$, which contains the indices of all the users associated with BS $k$ at slot $n$. 

\subsection{Adaptive Video Streaming Model}

In \ac{AS} over HTTP, the video content is divided into a sequence of small HTTP-based file segments. Video segments are then pre-encoded in multiple versions, each with a specific video bitrate and resolution or `quality level' \cite{oyman2012:QoE_HTTP}. Higher quality segments will be larger in size but represent similar playback durations. We denote the segment quality levels by $l\nabla\mathcal{Q}$, where $\mathcal{Q}=\{1,2,\ldots,q_\tn{max}\}$, and $q_\tn{max}$ is the maximum quality level. The function $f^Q_\tn{rate}(\cdot)$ maps the quality level to the corresponding bitrate. Higher segment qualities will require higher bitrates for successful reception, and therefore $f^Q_\tn{rate}(\cdot)$ is an increasing function of $l$. 
To assign the quality level of each user segment, we define the binary decision matrix $\mathbf{q}=(q_{i,s,l}\in\{0,1\}: {i\in\mathcal{M}},{s=\{1,2,\ldots,S\}},{l\in\mathcal{Q}})$. If there are three quality levels, and $q_{i,s,1}=1$, then user $i$ will receive segment $s$ at quality level $1$; and the remaining quality level indices are zero, i.e. $q_{i,s,2}=0$ and $q_{i,s,3}=0$. Therefore, to ensure that only one level is selected $\sum_{l=1}^{q_\tn{max}} q_{i,s,l}=1\,\,\forall\,i,s$.

\subsection{BS Power Consumption Model}
The \ac{BS} downlink power consumption is based on the linear load dependent power model \cite{C.Desset:DTXPowerModel},\cite{Earth}, where power is proportional to the \ac{BS} load, with a fixed power required at minimum load. For \ac{BS} $k$ at slot $n$, this can be represented as:
\begin{equation}
\label{eq:PowerModel}
p_{k,n}=\begin{cases}
    P_0 + (P_m-P_0)\, \tn{BS}_{k,n}^\tn{load}, & 0< \tn{BS}_{k,n}^\tn{load}\leq 1,\\
    P_\tn{sleep}, &  \tn{BS}_{k,n}^\tn{load}=0,
  \end{cases}
\end{equation}
where $P_m$ and $P_0$ are the power consumption at the maximum and minimum non-zero load, and \ac{BS} load is computed as $\tn{BS}_{k,n}^\tn{load}=\sum_{i \in \mathcal{U}_{k,n}}x_{i,n}$. When there is no load, the BS can enter a sleep mode which consumes $P_\tn{sleep}\,[W]$. Advanced BS hardware allows $P_\tn{sleep}$ to be a fraction of $P_0$, or even zero, thus, allowing a complete BS switch-off \cite{C.Desset:DTXPowerModel}, so we assume $P_\tn{sleep}=0$. \acp{BS} entering this deep sleep mode are required to remain off for at least $n_\tn{off}$ time slots to allow sufficient time before a wake-up is possible. We denote the BS power per slot matrix by $\mathbf{p}=(p_{k,n} \in [0,P_m]: {k\in\mathcal{K}},{n \in\mathcal{N}})$, and the BS on/off binary decision matrix by $\mathbf{b}=(b_{k,n} \in \{0,1\}: {k\in\mathcal{K}},{n \in\mathcal{N}})$.

\begin{figure*}
	\noindent
	\centering
	\includegraphics[trim=0.0cm 0cm 0cm 0cm, clip=true, width=120mm]{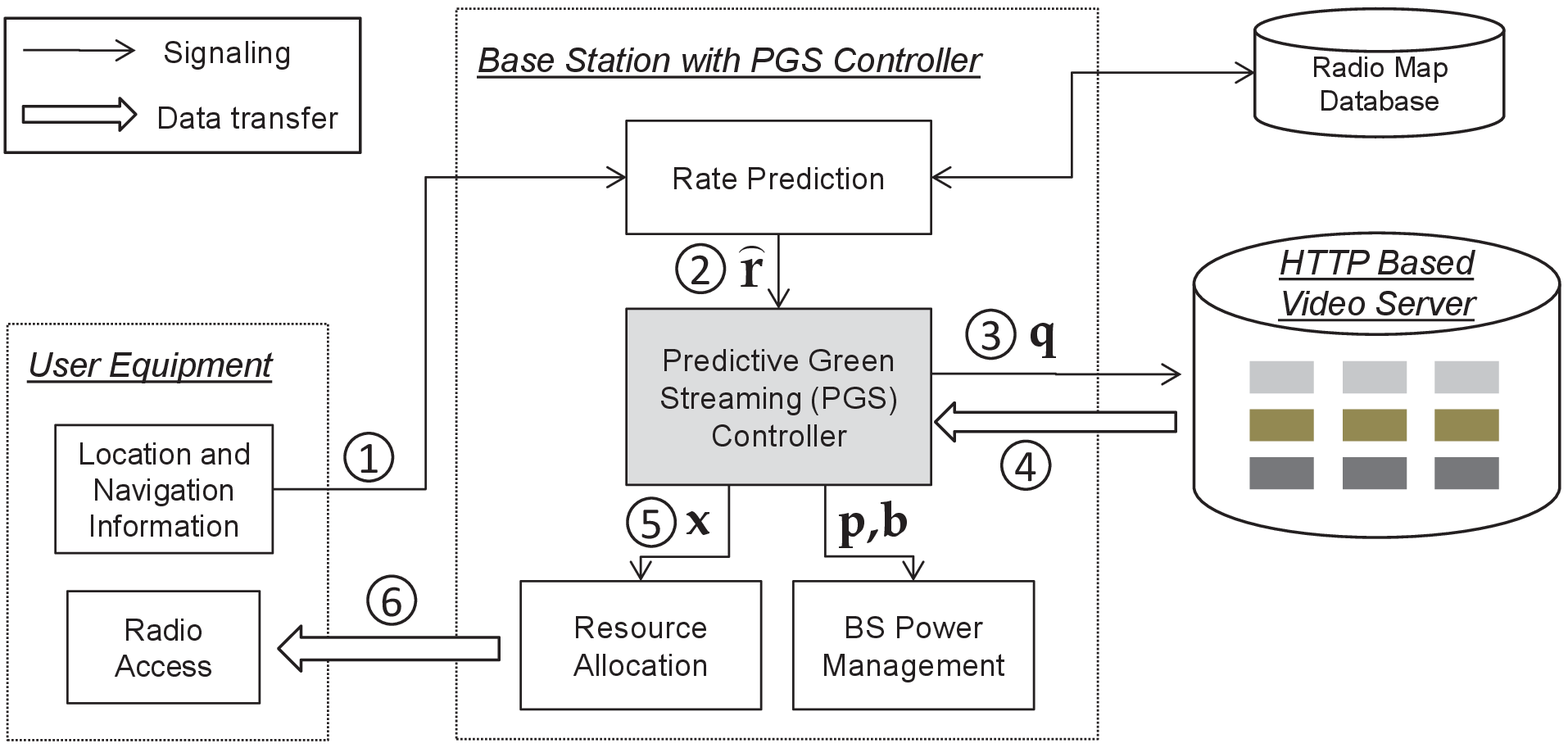} 
	\caption{\acl{PGS} Operation.}
	\label{fig:PGS_Framework}
\end{figure*}
\section{Predictive Green Streaming Framework}
\label{sec:PGS-MILP}
In this section, we present the \acl{PGS} framework that leverages rate prediction to minimize BS power consumption and transmission time of adaptive video streams. As opposed to live streaming, stored videos can be strategically delivered ahead of time and cached at the user equipment, after which transmission can be momentarily suspended while the user consumes the buffer. The essence of \ac{PGS} is that a \textit{long-term} resource allocation plan is made for each user by exploiting its individual rate predictions. By allocating over a time horizon, PGS can plan to grant more resources to users at their respective high data rates to prebuffer content efficiently, and reduce transmission energy.

\subsection{System Overview}
In \fref{fig:PGS_Framework} we present the considered architecture for HTTP-based \ac{AS} in the wireless network, and outline the required steps in \ac{PGS} to conceptualize its operation. First, we assume that user location and navigation information is provided to the \ac{BS}, which determines the future rates $\hat{\mathbf{r}}$ users will experience by consulting a radio map database. Since our focus is to  develop the predictive \ac{AS} transmission schemes, we assume $\hat{\mathbf{r}}$ is provided to the \ac{PGS} controller, and thereafter investigate the effect of prediction errors in \sref{sec:Sim}. The \ac{PGS} framework uses $\mathbf{\hat r}$ defined over a time horizon, to minimize power consumption while achieving a target video quality level with no video stalls. To do so, it jointly determines the optimal (i) user rate allocations $\mathbf{x}$, (ii) video segment qualities $\mathbf{q}$, (iii) BS transmit powers $\mathbf{p}$, and (iv) the BS on/off statuses $\mathbf{b}$. The required segments, as specified in $\mathbf{q}$, are requested from the HTTP-based video servers. These segments are then delivered to users over time slots in accordance with the determined rate allocation plan in $\mathbf{x}$. The \ac{PGS} controller also determines the deep sleep schedule of the BSs that minimizes power consumption without violating user requirements through the optimization variable $\mathbf{b}$, which is passed onto the BS power management unit. 

It is worth noting that currently \ac{DASH} relies on the client to signal the requested quality levels to the content server \cite{oyman2012:QoE_HTTP}. Therefore, the proposed in-network PGS approach requires some modifications to traditional DASH operation. However, there are current efforts towards enabling forms of network-assistance in DASH, under the MPEG Server and Network assisted DASH (SAND) operation \cite{SAND-DASH:MPEG}. 

We formulate two objectives of PGS as \acp{MILP} to provide benchmark solutions for performance evaluation. 
The first objective PGS-MinPower, minimizes total BS power consumption, where BSs can enter deep sleep,  under the constraint that no users experience video stalling. The \ac{MILP} formulation is non-trivial due to the tight coupling between the large number of optimization variables.
We then present PGS-MinAir with the objective of minimizing transmission airtime. However, in PGS-MinAir, BS turn-off is not enabled, and can be considered a special case of PGS-MinPower.
To formulate these problems, several constraints have to be considered, which can be classified into  1) user requirement constraints, and 2) \ac{BS} operation constraints.

\subsection{User Constraints}
\subsubsection{Rate Allocation for Smooth Streaming}

\begin{figure}[!t]
	\noindent
	\centering
	\includegraphics[trim=0cm 0cm 0cm 0cm, clip=true, width=80mm]{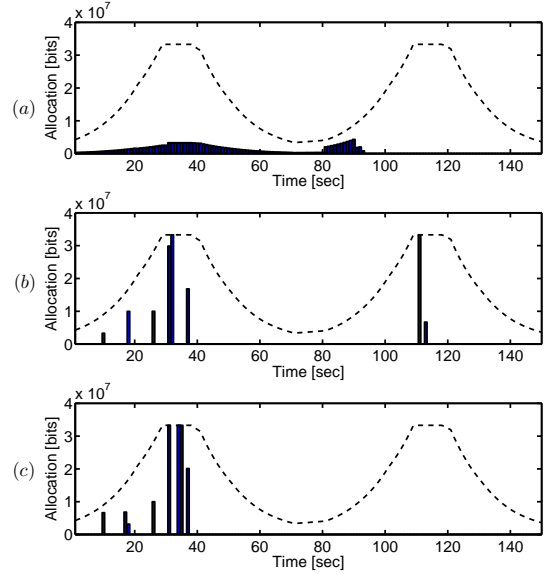} 
	\caption{Sample user allocation with time illustrating power minimization. (a) In traditional allocation, airtime is divided among users with per slot fairness considerations. (b) In PGS-MinAir, a user is pre-allocated video segments when the rate is high, to avoid inefficient allocation during low rate periods. (c) PGS-MinPower is similar to (b) with the additional objective of grouping user allocations to allow \acp{BS} to subsequently turn off.}
	\label{fig:UserAlloc_EE}
\end{figure}

Consider a user streaming a stored video at the \textit{maximum} quality level $q_\tn{max}$. For the video to playback without interruptions, $ f^Q_\tn{rate}(q_\tn{max})$ bits are required per second. Alternatively, a bulk of video content can be transmitted at once and buffered at the user's device, after which transmission can be suspended momentarily without causing video stalls. Therefore, we are interested in the cumulative video content stored at the user's device, which is given by $\sum_{n'=1}^n x_{i,n'}\,\hat{r}_{i,n'}$ at time slot $n$.
With the knowledge of $\mathbf{\hat r}$, an allocation plan can be made that grants minimum resources when the channel conditions are low, and prebuffers as much content as possible when conditions are high. This will reduce transmission time, thereby reducing BS load and saving power. 
To illustrate the idea, \fref{fig:UserAlloc_EE}(a) and (b) depict the difference between traditional allocation and the aforementioned predictive scheme where allocation is avoided during poor channel conditions. 


The joint relationship between the cumulative allocated rate and segment quality selection that ensures smooth playback in \ac{AS} is captured in the constraint
\begin{equation}
\label{eq:NoFreeze}
\tau_\tn{seg}\sum_{s'=1}^{s} \sum_{l=1}^{q_\tn{max}} q_{i,s',l}\, f^Q_\tn{rate}(l) \leq \sum_{n'=1}^{sN_\tn{seg}} x_{i,n'}\,\hat{r}_{i,n'} \,\,\,\, \forall\,i,\forall s, 
\end{equation}
\begin{equation}
\label{eq:OneLevel}
\sum_{l=1}^{q_\tn{max}} q_{i,s,l}=1, \qquad \forall\,i\in \mathcal{M},\forall s\in \{1,2,\cdots, S\}, 
\end{equation}
where \eref{eq:OneLevel} ensures that one quality level is selected. The \ac{RHS} of \eref{eq:NoFreeze} denotes the cumulative bits allocated to user $i$ at the slots corresponding to the end of each segment, whereas the \ac{LHS} expresses the cumulative bits \textit{required} to download up to $s$ video segments at the quality levels specified by $q_{i,s,l}$. For uninterrupted playback, an arbitrary segment $s$ must be completely downloaded, by time slot $sN_\tn{seg}$. Note that \eref{eq:NoFreeze} allows a trade-off between video quality and required airtime, while ensuring smooth playback. 

\subsubsection{Target Quality}
If $l_\tn{req} \in \{1,\cdots,q_\tn{max}\}$ denotes the desired average quality level for each user, then  
\begin{equation}
\label{eq:TargetQuality}
\sum_{s=1}^{S}\sum_{l=1}^{q_\tn{max}} q_{i,s,l} \geq l_\tn{req}S,\qquad \forall\,i\in \mathcal{M},
\end{equation}
represents the average user quality constraint. 
\subsubsection{User Buffer Limit}
In addition to the key constraints in \eref{eq:NoFreeze} and \eref{eq:TargetQuality}, a limit can also be imposed on the number of bits that can be pre-buffered at the user's device. This may be due to the video client and network policy, or device limitations. If $L_i$ denotes the limit for user $i$, then we have the constraint
\begin{equation}
\label{eq:BuffLimit}
\begin{aligned}
&\sum_{n'=1}^n x_{i,n'}\,\hat{r}_{i,n'} -\tau_\tn{seg}\sum_{s'=1}^{\lfloor n/N_\tn{seg}\rfloor}\sum_{l=1}^{q_\tn{max}} q_{i,s',l}\, f^Q_\tn{rate}(l)&\\
&-(n\tn{mod}({N_\tn{seg}})) \sum_{l=1}^{q_\tn{max}} q_{i,\lceil n/N_\tn{seg}\rceil,l}\, f^Q_\tn{rate}(l) \leq L_i, \quad \forall\,i,\forall n.&
\end{aligned}
\end{equation}
The \ac{LHS} of \eref{eq:BuffLimit} determines the difference between the cumulative allocated bits and the bits required for smooth playback at every slot $n$, and therefore denotes the buffered bits. The second term on the \ac{LHS} accounts for the bits of previously played video segments, and the third term represents the portion of the current segment that has been played. 

\begin{figure*}[!t]
\noindent
\centering
	\subfigure[Traditional non-predictive operation.]{
	\includegraphics[width=58mm]{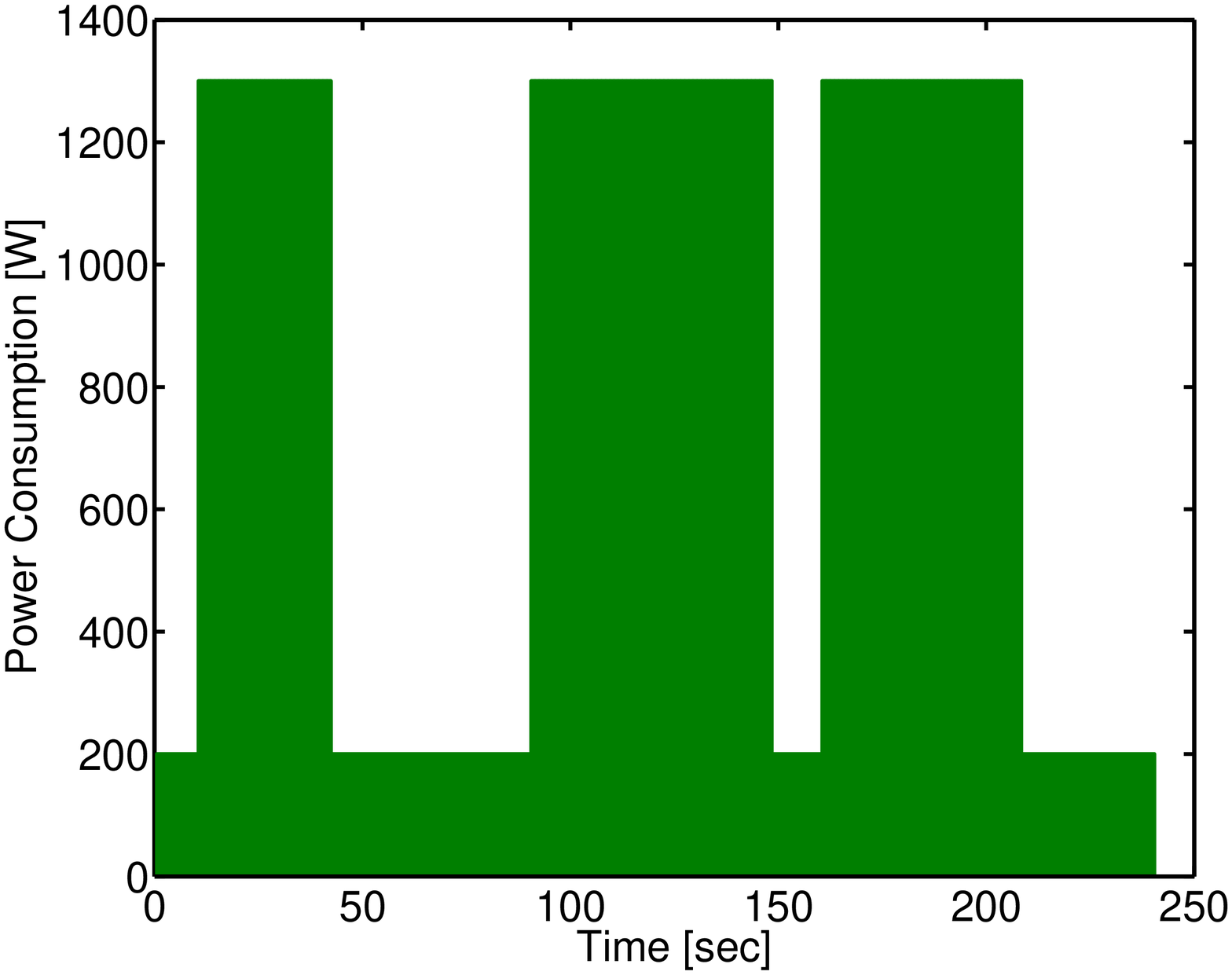}\label{fig:BSPowerES}}
	\subfigure[PGS-Airtime Minimization.]{
	\includegraphics[width=58mm]{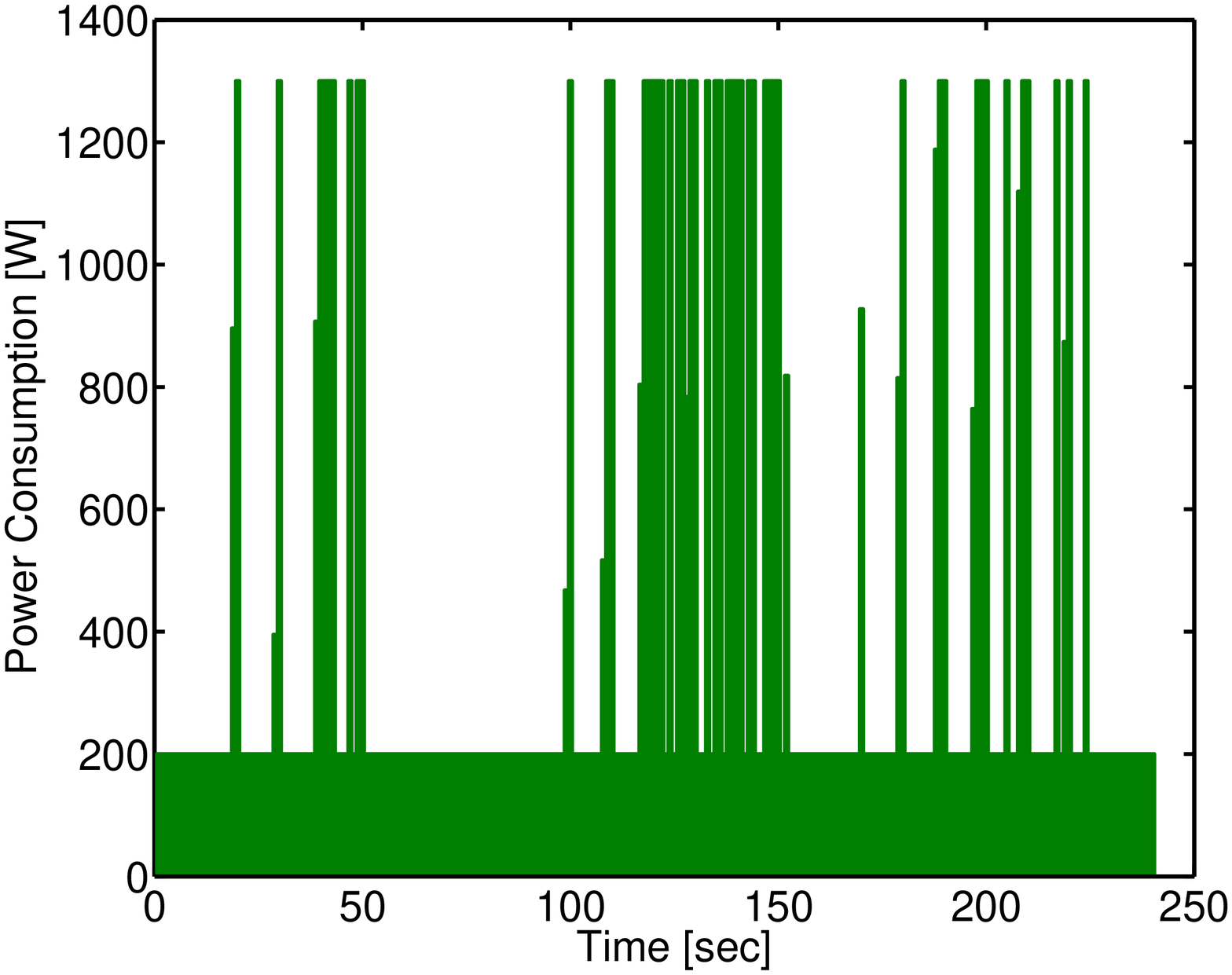}\label{fig:BSPowerMinAir}}
	\subfigure[PGS-Power Minimization.]{ 
	\includegraphics[width=58mm]{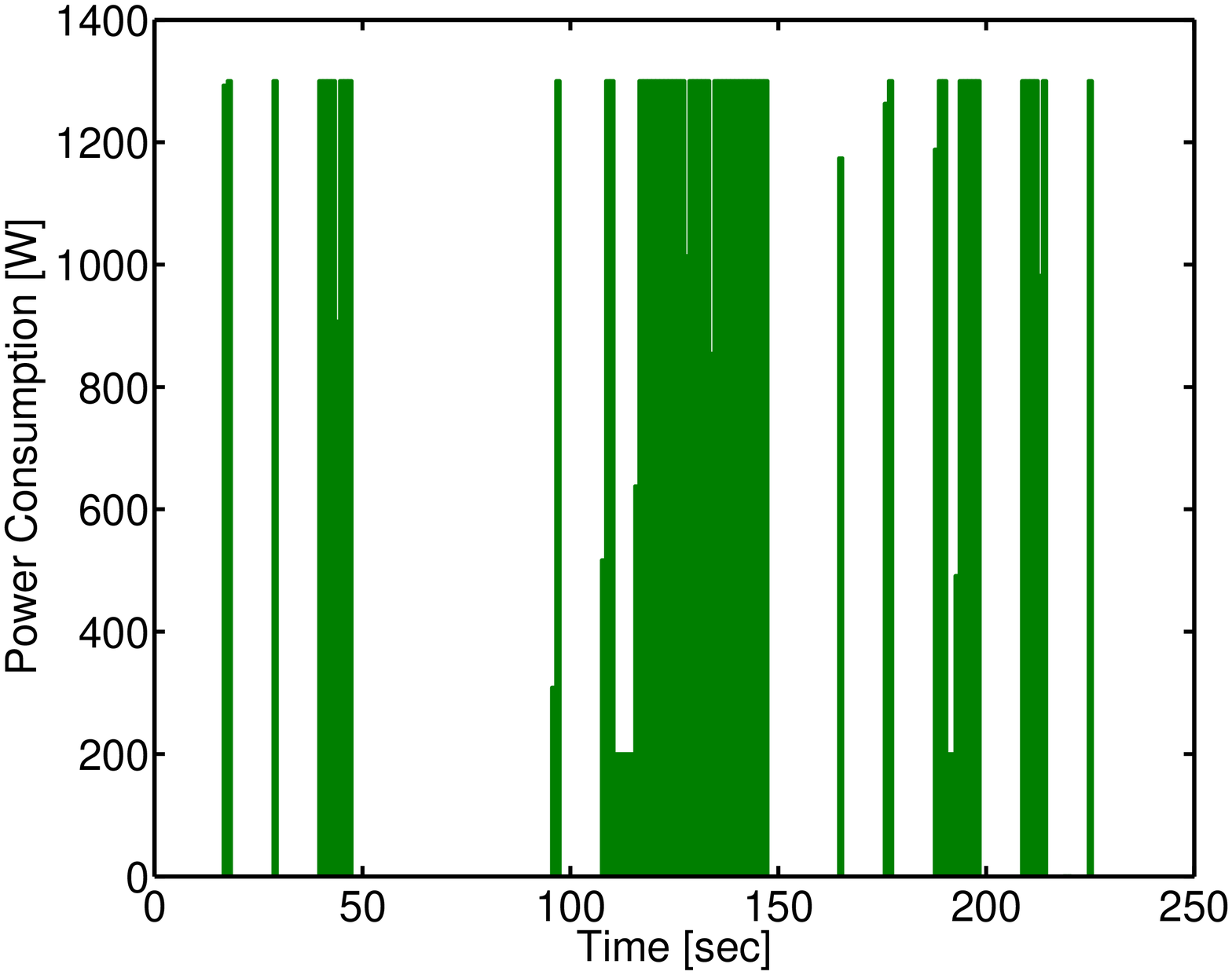}\label{fig:BSPowerMinPower}}
	\caption{Sample \ac{BS} power consumption over time, where $P_0=200\,$W and $P_m=1300\,$W. (a) In traditional operation, BS airtime is fully utilized when there are users present. (b) With PGS-MinAir, BS airtime is minimized by opportunistic allocations using rate predictions as shown in \fref{fig:UserAlloc_EE}(b). In (c) PGS-MinPower allows the BS to enter deep sleep modes and conserve more power. To do so, user allocations are grouped together as illustrated in  \fref{fig:UserAlloc_EE}(c).}
	\label{fig:BSPower}
\end{figure*}
\subsection{BS Constraints}
\subsubsection{BS Resource Limit}
The BS resource constraint limits the sum of the user airtime fractions to unity, i.e.:
\begin{equation}
\label{eq:BSX}
\sum_{i \in \mathcal{U}_{k,n}} x_{i,n} \leq 1,\qquad \forall\,k\in \mathcal{K},\,\forall n\in \mathcal{N}.
\end{equation}
This constraint is applied at each BS, where the summation is over all users $i$ associated with BS $k$ at slot $n$. 
\subsubsection{BS Slot Power Consumption} According to the BS power model of \eref{eq:PowerModel}, the power consumed by each BS is dependent on (i) the total user airtime, and (ii) whether the BS is kept on or switched off. This is expressed by the following constraint
\begin{equation}
\label{eq:PowerVarDef}
(P_\tn{m}-P_\tn{0})\sum_{i \in \mathcal{U}_{k,n}} x_{i,n} - p_{k,n} + P_\tn{0}\,b_{k,n}=0,\,\,\,\forall\,k\in \mathcal{K},\,\forall n\in \mathcal{N},
\end{equation}
where the binary decision variable $b_{k,n}$ is multiplied by $P_0$ to produce zero power consumption when the BS is off. 
\subsubsection{BS On Constraint} To enforce a BS to be on if there is any load, we apply the following constraint
\begin{equation}
\label{eq:BSONDef}
\sum_{i \in \mathcal{U}_{k,n}} x_{i,n} - b_{k,n}\leq 0,\,\,\,\forall\,k\in \mathcal{K},\,\forall n\in \mathcal{N}.
\end{equation}
\subsubsection{BS Off Indicator}
In order to monitor when a BS is turned off, we introduce an indicator variable $I_{k,n}$ which is equal to 1 only when a BS is turned off. This is achieved by
\begin{equation}
\label{eq:BSOFFDef}
- b_{k,n-1}+ b_{k,n} + I_{k,n} = 0,\,\,\,\forall\,k\in \mathcal{K},\,\forall n\in \mathcal{N},
\end{equation}
where $b_{k,0}=0 \,\,\forall\,k$. On the other hand, when a BS is switched on, $I_{k,n}=-1$, and if there is no change $I_{k,n}=0$. The value of this indicator is used by the following constraint to ensure that a BS remains off for a minimum number of $n_\tn{off}$ slots.
\subsubsection{Minimum Off Time} To model the minimum off duration, we restrain the BS from turning on for $n_\tn{off}$ slots, once it has been turned off. This can be achieved by
\begin{equation}
\label{eq:BSOFFLim}
b_{k,n+c}+ I_{k,n} \leq 1,\,\,\,\forall\,k\in \mathcal{K},\,\forall n\in \mathcal{N},\,\forall c,
\end{equation}
where $c={1,\cdots,n_\tn{off}},$ and $n+c<=N$. Equation \eref{eq:BSOFFLim} ensures that if the indicator of the previous time slot is 1, then $b_{k,n+c}$ will have to remain equal to zero for $n_\tn{off}$ slots. This is controlled through the variable $c$ that generates $n_\tn{off}$ constraints to define the on/off status of the upcoming $n_\tn{off}$ slots, for every $n$. If on the other had, the indicator is not 1, then $b_{k,n+c}$ can take on any value. 
\subsection{PGS-MinPower MILP Problem Definition}
Considering the previously discussed constraints, the PGS-MinPower problem can be formulated as the following MILP:
\begin{equation}
\label{eq:PGS-MinPower}
\underset{\mathbf{x},\mathbf{q},\mathbf{p},\mathbf{b}}{\text{minimize}} \quad\,\,
  \sum_{k=1}^{K} \sum_{n=1}^{N} p_{k,n}
\end{equation}
\begin{equation*}
\begin{aligned}
& \text{subject to:}
& & \tn{Constraints}\,\,\eref{eq:NoFreeze} \tn{ to } 
\eref{eq:BSOFFLim}
\\
& & & 0 \leq x_{i,n} \leq 1,\,\,\,\forall\,i\in\mathcal{M},\forall n\in \mathcal{N},\\
& & & q_{i,s,l}\in \{0,1\},\,\,\,\forall i\in\mathcal{M},\forall s\in \mathcal{S},\forall l\in \mathcal{Q},\\  
& & & 0 \leq p_{k,n} \leq P_m,\,\,\,\forall\,k\in\mathcal{K},\forall n\in \mathcal{N},\\
& & & b_{k,n} \in \{0,1\},\,\,\,\forall\,k\in\mathcal{K},\forall n\in \mathcal{N}.\\
\end{aligned}
\end{equation*}

Note that although \eref{eq:PGS-MinPower} provides the optimal joint allocations of all the decision variables, it is computationally intractable to solve large instances of PGS-MinPower due to the large number of MN+MS+2KN decision variables, and the coupling between them. Further, considerable memory is needed as the resulting constraint matrix has a size of M+MN+2MS+5KN, which can be very large. The duration of the lookahead window impacts both N and S, and therefore, the complexity of \eref{eq:PGS-MinPower} depends primarily on the prediction window duration. 

It is worth noting that overhead may be introduced when  turning BSs off/on. This may be accounted for by increasing the value of $n_\tn{off}$ to prevent frequent, short sleeps. Another way to directly incorporate this overhead is through the BS Off Indicator variable $I_{k,n}$ defined in \eref{eq:BSOFFDef}. This can be achieved by adding another power consumption term  to the objective in \eref{eq:PGS-MinPower}, which sums over $I_{k,n}$ multiplied by a constant that denotes the power consumption of a single on/off switch. The PGS solution will then minimize the total power consumed while accounting for the overhead of the deep sleep switches.

\subsection{PGS-MinAir MILP Problem Definition}
The PGS-MinAir problem considers the case where BSs cannot be switched off into deep sleep modes, for example due to other types of traffic in the network. PGS-MinAir can be formulated by setting the BS on/off decision variable to 1 and excluding constraints \eref{eq:BSONDef} to \eref{eq:BSOFFLim} in \eref{eq:PGS-MinPower}. However, a more compact formulation can also exclude the BS power $p_{k,n}$ variables, and airtime can be minimized directly through user allocations $x_{i,n}$. This is possible since BS power is proportional to airtime in the linear BS power model of \eref{eq:PowerModel}. Therefore, the PGS-Airtime problem can be formulated as
\begin{equation}
\label{eq:PGS-MinAir}
\underset{\mathbf{x},\mathbf{q}}{\text{minimize}} \quad\,\,
  \sum_{i=1}^M \sum_{n=1}^{N} x_{i,n}
\end{equation}
\begin{equation*}
\begin{aligned}
& \text{subject to:}
& & \tn{Constraints}\,\,\eref{eq:NoFreeze} \tn{ to } 
\eref{eq:BSX},
\\
& & & 0 \leq x_{i,n} \leq 1,\,\,\,\forall\,i\in\mathcal{M},\forall n\in \mathcal{N},\\
& & & q_{i,s,l}\in \{0,1\},\,\,\,\forall i\in\mathcal{M},\forall s\in \mathcal{S},\forall l\in \mathcal{Q}.\\  
\end{aligned}
\end{equation*}
\fref{fig:BSPower} depicts an example of the resulting BS power consumption plan for \ac{PGS} versus a traditional scheme, where BS airtime is shared equally among video streaming users. In the scenario considered, vehicular users arrive at the BS in three consecutive groups. In \fref{fig:BSPowerES}, as long as users are present, BS airtime is completely utilized. However, in \fref{fig:BSPowerMinAir} and \fref{fig:BSPowerMinPower}, \ac{PGS} allows the BS to transmit in a spectrally efficient way without violating user streaming requirements. Note that while \ac{PGS}-MinAir minimizes total transmit time, \ac{PGS}-MinPower in \fref{fig:BSPowerMinPower} is able to strike the optimal tradeoff between serving users when their individual rates are high, and grouping user transmissions together (even if not at their respective best rates) to generate blocks of sleep time. This comes at the cost of an increased complexity as observed in the \ac{PGS}-MinPower formulation, where the optimization variables are tightly coupled.
However, at high load, the power saving gains of PGS-MinPower over PGS-MinAir will decrease and eventually converge to PGS-MinAir. This is due to the decreased ability to generate silent space long enough for a BS switch off. We discuss more details of the tradeoffs involved in the numerical results of \sref{sec:Sim}. Before that, we present a polynomial-time solution of the PGS the problem, which achieves close to optimal results.

\section{Multi-stage \ac{PGS} Solution}
\label{sec:PGS-Heuristic}

\begin{figure}[!b]
	\noindent
	\centering
	\includegraphics[trim=0.0cm 0cm 0cm 0cm, clip=true, width=65mm]{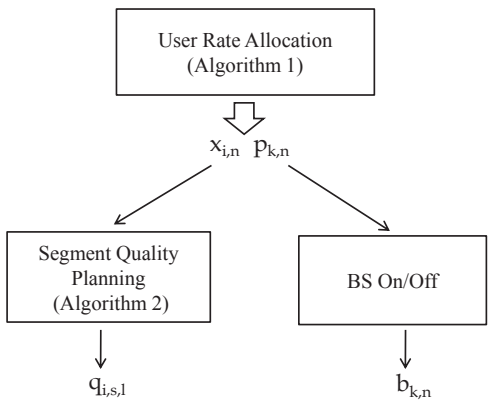} 
	\caption{The Proposed Multi-Stage \ac{PGS} Solution.}
	\label{fig:PGS_Steps}
\end{figure}

In this section, we develop a multi-stage approach to solve the \ac{PGS} \acp{MILP} presented in \sref{sec:PGS-MILP}. \fref{fig:PGS_Steps} outlines the steps involved. The core stage is a user rate allocation algorithm that assigns BS airtime to users over the lookahead window, thereby solving $\mathbf{x}$ and $\mathbf{p}$. Thereafter, segment qualities are explicitly planned for each user based on the allocated bits, and BS on/off statuses are determined from the resulting idle time in $\mathbf{p}$. This decoupling is based on the intuition that an efficient rate allocation scheme exploiting rate predictions will provide power savings while satisfying user quality needs. Before discussing each stage, we introduce the following definitions:
\begin{itemize}
\item Cumulative demand $D_{i,s}$: the total number of bits required by user $i$ to stream the first $s$ segments. For a given target quality level $l_\tn{req}$, $D_{i,s}=s\, f^Q_\tn{rate}(l_\tn{req})\,[\tn{bits}], \,\,\forall\,i $.
\item Cumulative rate allocation $R_{i,n}$: the total rate allocated to user $i$ by time slot $n$, i.e. ${R_{i,n}=\sum_{n'=1}^n x_{i,n'} \hat r_{i,n'}},\,\,\forall\,i$.
\item User rate percentile $\hat r^{y\%}_{i,n}$: the $y^\tn{th}$ percentile of the future user rate, i.e. computed over $\hat r_{i,n\leq n' \leq N}$, for each user.
\end{itemize}

\subsection{Rate Allocation}
The \ac{RA} strategy is to divide the problem into a series of allocation subproblems performed at the start of each segment. The idea of this decomposition is to minimize airtime while focusing on satisfying the streaming constraint in \eref{eq:NoFreeze}, that is performed for each segment. If $\mathcal{N}^s$ denotes the set of slots comprising segment $s$, then $\mathcal{N}^s = \{(s-1)N_\tn{seg}+1,(s-1)N_\tn{seg}+2,\cdots, sN_\tn{seg}\}$, and allocation is made incrementally for each $\mathcal{N}^s$. Each allocation is further divided into two steps 1) airtime minimization, and 2) opportunistic prebuffering. In the first step, users that do not have enough content prebuffered to stream the upcoming segment at the target quality level are prioritized, and their demands are fulfilled with the minimum airtime. In the second step, users that have exceptionally good channel conditions are granted excess airtime to prebuffer future video content. This will reduce the airtime required later to download upcoming segments. Next, we discuss the details of each step.

\subsubsection{Airtime Minimization}
\label{sec:AirMin}
At the start of segment $s$, each BS $k$ determines the set of priority users $\mathcal{P}_{k,s}$ that have insufficient cumulative allocation to play the upcoming video segment at the target quality level. The set $\mathcal{P}_{k,s}$ will, therefore, not include users that have prebuffered segments. The required rate allocation $r^{\mathcal{P}}_{i}$ for user $i$ can then be computed as
\begin{equation}
\label{eq:Def_i}
r^{\mathcal{P}}_{i}= D_{i,s} - R_{i,sN_\tn{seg}},\quad\forall i \in \mathcal{P}_{k,s},
\end{equation}
where $R_{i,0}=0\,\,\forall i$. To serve the users with these rate requirements, using the minimum BS airtime, we need to solve the optimization problem 
\begin{equation}
\label{eq:PGS-MinAirSeg}
\underset{\mathbf{x},\mathbf{Y}}{\text{minimize}} \quad\,\,
\sum_{n \in \mathcal{N}^s} 
\sum_{i \in {\mathcal{P}_{k,s}}} x_{i,n} + \beta \sum_{i \in {\mathcal{P}_{k,s}}} Y_{i}
\end{equation}
\begin{equation*}
\begin{aligned}
&\text{subject to:}&&-\sum_{i \in \mathcal{P}_{k,s}} x_{i,n}\hat r_{i,n}-Y_i\leq -r^{\mathcal{P}}_{i},\quad \qquad\forall n\in \mathcal{N}^s,\\
&&&\sum_{i \in \mathcal{P}_{k,s}} x_{i,n} \leq 1, \quad \qquad \qquad \qquad \qquad\forall n\in \mathcal{N}^s,\\
&&&0 \leq x_{i,n} \leq 1, \qquad \qquad \qquad \forall\,i\in \mathcal{P}_{k,s},n\in \mathcal{N}^s,\\
&&&0 \leq Y_{i}, \qquad \qquad \qquad \qquad \qquad \qquad\,\forall\,i\in \mathcal{P}_{k,s}.\\
\end{aligned}
\end{equation*}
The variables $Y_i$ are introduced to capture any unfulfilled rate allocation, when it is not possible to satisfy all user requirements. As satisfying users with the target quality level has a higher precedence over saving airtime, the weight parameter $\beta>1$. Generally, at low to moderate loads (where there are potential power savings), \eref{eq:PGS-MinAirSeg} will yield $Y_i=0$, and quality requirements will be met with minimum BS airtime. 

Note that the problem in \eref{eq:PGS-MinAirSeg} has a linear objective function with linear constraints and is therefore a \ac{LP}, which can generally be solved efficiently, even with the widely-used Simplex algorithm \cite{LPTimeComplexitySimplex}. Further, the problem dimension is incomparable to the optimal \ac{PGS} \ac{MILP} formulations, and therefore provides significant computational and memory requirement gains.

Alternatively, to avoid the requirement of BSs being equipped with \ac{LP} solvers, we present the following simple algorithm to solve  \eref{eq:PGS-MinAirSeg}. First, the set $\mathcal{P}_{k,s}$ is sorted in descending order of user requirements $r^{\mathcal{P}}_{i}$. Then, each user $i\in \mathcal P_{k,n}$ is selected in sequence to transmit at the time slot $n^* \in \mathcal{N}^s$, that has the largest predicted rate for that user, i.e.
\begin{equation}
\label{eq:maxUserSlot}
n^*=\arg \max_n \, \hat r_{i,n}, \quad \forall n \in \mathcal{N}^s.
\end{equation}
Note that if $\tn{BS}^\tn{air}_{k,n}$ denotes the airtime available in BS $k$ at slot $n$, then the search in \eref{eq:maxUserSlot} will exclude slots where $\tn{BS}^\tn{air}_{k,n}=0$. The airtime allocated to the user is then $x_{i,n^*}=\hat r_{i,n^*}/r^{\mathcal{P}}_{i}$, and the remaining BS airtime is updated to $\tn{BS}^\tn{air}_{k,n^*}=\tn{BS}^\tn{air}_{k,n^*}-x_{i,n^*}$.
After iterating over all $i\in \mathcal P_{k,n}$, we update $R_{i,n},\mathcal{P}_{k,s}$ and $r^{\mathcal{P}}_{i}$, and the process is repeated until either $\mathcal{P}_{k,s}=\phi$ or there is no remaining BS airtime for $n \in\mathcal{N}^s$. This procedure is outlined in lines 6-14 in \aref{alg:RateAlloc}, and numerical results in \sref{sec:Sim} indicate that it provides almost identical results to the \ac{LP} of \eref{eq:PGS-MinAirSeg}.
\begin{algorithm}[!t]
\caption{User Rate Allocation Algorithm}
\label{alg:RateAlloc}
\begin{algorithmic}[1]
\REQUIRE $\hat{r}_{i,n}, \mathcal{U}_{k,n},D_{i,s},K,M,N,N_\tn{seg}$
\STATE Initialize $x_{i,n},R_{i,n}=0 \quad \forall\,i,n=1,2,..N$
\FORALL {$y \in\{65,70,\cdots,95\}$}
\STATE Reset $x_{i,n}=0,\,\tn{BS}^\tn{air}_{k,n}=1 \quad \forall i,k,n.$
\FORALL {segments $s$}
    \FORALL {base stations $k$}
		\STATE Find set of priority users $\mathcal{P}_{k,s}$, and compute $r^\mathcal{P}_i$ as in \eref{eq:Def_i}. Sort $\mathcal{P}_{k,s}$ in descending order of $r^\mathcal{P}_i$.
		\WHILE {$\mathcal{P}_{k,s}\neq \phi$ \AND $\sum_{n \in \mathcal{N}^s}\tn{BS}^\tn{air}_{k,n}>0$}
		\FORALL {users $i \in \mathcal{P}_{k,s}$}
			\STATE	Find slot $n^*$ with the largest rate as in \eref{eq:maxUserSlot}. 
			\STATE Set $x_{i,n^*}= \hat r_{i,n^*}/r^\mathcal{P}_i$.
			\STATE Set $\tn{BS}^\tn{air}_{k,n^*}=\tn{BS}^\tn{air}_{k,n^*}-x_{i,n^*}$
			\ENDFOR
			\STATE Recompute $R_{i,n},\mathcal{P}_{k,s}$ and$ r^\mathcal{P}_i$.
		\ENDWHILE
  		\FORALL {slots $n \in \mathcal{N}^s$}
		\STATE Find user $i^*$ with the largest $\hat r_{i,n} \quad \forall i \in \mathcal{U}_{k,n}.$
		\STATE If $\hat r_{i^*,n}>r^y_{i^*}$, then  $x_{i^*,n}=x_{i^*,n}+\tn{BS}^\tn{air}_{k,n}$.
		\ENDFOR
	\ENDFOR
\ENDFOR
\STATE Calculate $p_{k,n}$ using \eref{eq:PowerModel}, where $\tn{BS}_{k,n}^\tn{load}=1-\tn{BS}^\tn{air}_{k,n}$.
\STATE Calculate $P^y_\tn{Net}=\sum_{k=1}^K\sum_{n=1}^N p_{k,n}$ for this iteration.
\ENDFOR
\STATE Determine $y^*$ that produces the minimum $P^y_\tn{Net}$.
\RETURN $\mathbf{x},\mathbf{p}$
\end{algorithmic}
\end{algorithm}

\subsubsection{Opportunistic Prebuffering}
While the airtime minimization stage provides users with their \textit{immediate} needs efficiently, it does not capitalize on granting users their \textit{future} content in advance when their rates are high. Implementing such prebuffering results in reduced overall airtime since bulk transmissions are made opportunistically in short time durations, and thereafter users are not served. However, the question remains: when is a good time to prebuffer content to a user? A simple rate threshold will not work well for cases where users have unequal rate distributions over $\mathcal{N}$. We, therefore, use the previously defined rate percentile $\hat r^{y\%}_{i,n}$ metric, as it provides each user with an independent threshold, derived from its own rate statistics. 
This is applied as follows: for each slot $n\in\mathcal{N}^s$, we first find the user $i^*$ with the largest rate, i.e. 
\begin{equation}
i^*=\arg \max_i \, \hat r_{i,n}\,\quad \forall i \in \mathcal{U}_{k,n}.
\end{equation}
This rate is then compared to the user's $y^\tn{th}$ rate percentile at $n$, and if  ${\hat r_{i^*,n}>r^{y\%}_{i^*,n}}$, the user is allocated the remaining BS airtime at that slot, and the user airtime is updated to $x_{i^*,n}=x_{i^*,n}+\tn{BS}^\tn{air}_{k,n}$. 
This completes the two steps of rate allocation performed $\forall n\in\mathcal{N}^s$. The procedure is then repeated by each BS, for each segment in sequence, as outlined in \aref{alg:RateAlloc}. The 
BS power consumption matrix $\mathbf{p}$ is then calculated using \eref{eq:PowerModel}, where $\tn{BS}_{k,n}^\tn{load}=1-\tn{BS}^\tn{air}_{k,n}$. \\
\textbf{Setting y}: The value of $y$ can affect the resulting power savings and is dependent on the current network load. At low load, a higher $y$ will cause users to only prebuffer at close to peak rates. This is more efficient, provided users do not thereafter fall short of their needs and request airtime before encountering another `peak'. On the other hand, when load is high, a lower value of $y$ is preferred to allow users to prebuffer more frequently, even if at moderate rates. 
Although intermediate values $y\in [70,80]$ provide a good tradeoff, the optimum value can be determined by iterating the procedure for different values and selecting the rate allocation $\mathbf{x}$ that gives the minimum power consumption.

\begin{algorithm}[!t]
\caption{Segment Quality Algorithm}
\label{alg:QAlg}
\begin{algorithmic}[1]
\REQUIRE $x_{i,n}, \hat{r}_{i,n},q_\tn{max},M,N,N_\tn{seg},S$
\STATE Initialize $q_{i,s,1}=1 \,\, \forall\,i,s$\,[lowest quality level]
	\FORALL {users $i$}
		\FORALL {segments $s$} %
   			\STATE Set current segment quality to the highest level\\ $l^*=q_\tn{max},\,\,q_{i,s,l}=1 $ \tn{for} $l=l^*,\,\,q_{i,s,l}=0\,\,\forall\,l\neq l^*$. 
			\WHILE{$l^*\geq 0$ \AND \eref{eq:NoFreeze} is violated for any $s$}
			\STATE Lower current segment quality, $l^*=l^*-1$,\\
			$q_{i,s,l}=1 $ \tn{for} $l=l^*,\,\,q_{i,s,l}=0\,\,\forall\,l\neq l^*$.
   			\ENDWHILE
   			\ENDFOR
   	\ENDFOR
 \RETURN $\mathbf{q}$
 \end{algorithmic}
 \end{algorithm}

\subsection{Segment Quality Algorithm}
After determining the rate allocation matrix $\mathbf{x}$ as specified in \aref{alg:RateAlloc}, the user segment quality levels are planned. 
The objective is to determine the segment quality plan that maximizes quality while providing smooth playback. The idea is to iterate over the segments in sequence and greedily maximize the current segment quality, while ensuring that the future segments can be streamed, at least, at the lowest quality level. On average, the quality levels will be equal to $l_\tn{req}$. This is achieved as follows: all the segments are first initialized to a quality level of 1. Then, at the start of each segment, the quality level is set to $q_\tn{max}$ and a check is made to ensure that constraint \eref{eq:NoFreeze} is satisfied for $s,s+1,\cdots S$. If this is not the case, the current segment quality is iteratively decremented, until the constraints are met or the quality level is zero. The complete procedure is outlined in \aref{alg:QAlg}.

Note that \aref{alg:QAlg} is applied to each user independently as the resource allocation has already been determined. A practical property of the algorithm is that it ensures users experience the highest quality level as soon as possible, and for the longest possible duration. This is not guaranteed by solving \eref{eq:PGS-MinPower} or \eref{eq:PGS-MinAir}, since when a mix of low and high quality segments are pre-buffered, they can be ordered arbitrarily while remaining equivalently optimal. 
Therefore, \aref{alg:QAlg} can also be used to `post-process' the optimal result of ${\mathbf{x}}$ in the \ac{MILP} solutions, to generate ${\mathbf{q}}$ solutions that favor `early' high quality streaming. 

\subsection{BS On/Off Switching}
To determine the BS on/off status we simply search each \ac{BS} for long 'silent' transmission durations over the lookahead window, where there is zero load.
This is accomplished by the following simple procedure: 1) determine the set of time slots $\mathcal{N}_\tn{On}$ where $p_{k,n}>P_0$, implying that the BS is on; 2) then determine the difference between the successive time slots in $\mathcal{N}_\tn{On}$. If this is larger than $n_\tn{off}$, it means that no transmission occurred for a duration long enough to turn the BS off for that period. A value of zero is subtracted from the first element of $\mathcal{N}_\tn{On}$ to account for the possibility of switching the BS off before the first start, and the last element of $\mathcal{N}_\tn{On}$ is subtracted from $N$ to check for a turn off possibility at the end.

This completes the multi-stage \ac{PGS} solution, which we refer to as PGS-MinPower-Alg. For the case where BSs cannot switch to deep sleep we do not apply the BS On/Off stage, and only airtime is minimized. This solution is denoted by PGS-MinAir-Alg. Finally, when implementing the \ac{LP} of \eref{eq:PGS-MinAirSeg}, the algorithm will be denoted by PGS-MinAir-AlgLP.

\subsection{Computational Complexity}
We first determine the complexity of each stage of the PGS multi-stage solution. The airtime minimization step of the rate allocation involves computing \eref{eq:Def_i} and sorting the set $\mathcal{P}_{k,n}$, which has a time complexity of $O(MN+M\log M)$. Then, rate allocation over the $N_\tn{seg}$ slots takes $O(MN_\tn{seg})$ time, leading to an overall complexity of $O(MN+M\log M+MN_\tn{seg})$ for this step. The subsequent prebuffering includes computing the future rate percentile stage and takes $O(N_\tn{seg}(M+N\log N))$ time. After accounting for $S$ segments for each user, we arrive at an overall complexity of $O(MN^2)$ for \aref{alg:RateAlloc}. In the segment quality algorithm, the core step is to evaluate constraint \eref{eq:NoFreeze}, which has as a time complexity $O(N+S)$ for a single user. This step is repeated at most $q_\tn{max}$ times when the constraint is violated, and repeated for each segment and each user. The resulting complexity order is $O(q_\tn{max}MS(N+S))$. In the worst case $S=N$, and $q_\tn{max}$ is typically less than $6$, which gives a worst case runtime of $O(MN^2)$. 
As the BS on/off procedure presented earlier has a complexity of $O(KN^2)$, this leads to overall runtime of $O((M+K)N^2)$.

\section{Numerical Results}
\label{sec:Sim}
In this section, we present numerical results that demonstrate the potential energy savings achieved by exploiting rate predictions in the \ac{PGS} framework. We also investigate the effects of prediction errors on the performance of the \ac{PGS} schemes. 
\subsection{Simulation Setup}
\begin{figure}[!t]
\noindent
\centering
\includegraphics[width=80mm]{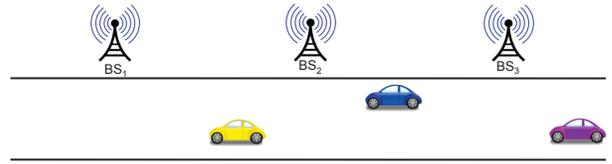}
\caption{Highway scenario with vehicular mobility.}
\label{fig:Highway}
\end{figure}

We consider two network set-ups. The first is a single cell with vehicles moving along a highway that crosses through the cell. And the second is a three \ac{BS} network, also along a highway, with an inter-BS distance of 1 km as shown in \fref{fig:Highway}. For realistic vehicular mobility we use the SUMO traffic simulator \cite{SUMO2011} to generate mobility traces with a flow of 1 vehicle per second. Vehicles arrive in groups of ten vehicles each, separated by 60 seconds. This creates the effect of vehicle grouping observed on highways. 

BS transmit power is $40$\,W, and bandwidth is $5$\,MHz. BS power consumption at minimum and maximum load is $200$\,W and  $1300$\,W respectively, as presented for macro BSs employing time duty-cycling in the power model of \cite{C.Desset:DTXPowerModel}. The minimum off time for a \ac{BS} is set to $10$\,s. The slot duration is $\tau=1$\,s and $T=240$\,s. We consider a video format with four quality levels of $\{0.25,0.5,0.75,1\}\,$Mbps, and a segment length $\tau_\tn{seg}=10$\,s. 
Gurobi 5.1 \cite{Gurobi} is used to solve the PGS \acp{MILP}, and Matlab was used as a simulation environment. 

We compare the performance of the \ac{PGS} schemes against two baseline approaches that do not exploit rate predictions. These reference schemes have two stages: rate allocation, followed by quality adaptation. Two rate allocation schemes are considered: \ac{ES} and \ac{RP}. In \ac{ES}, airtime is shared equally among the users at each time slot. If there are $N_{k,n}$ users associated with \ac{BS} $k$ at time $n$, then ${x_{i,n}=1/N_{k,n}\, \forall i \in \mathcal{U}_{k,n}}$, and the rate allocated to each is $\hat r_{i,n}/N_{k,n}$. The \ac{RP} allocator is designed to be more spectrally efficient but not fair to users. Here, airtime is assigned to each user in proportion to its achievable data-rate. Therefore, ${x_{i,n}=\hat r_{i,n}/\sum_{i\in \mathcal{U}_{k,n}} \hat r_{i,n}}$. Segment quality is then adapted based on the allocated rate at the start of the current segment, and the highest supportable level is selected. These approaches are referred to as ES-AdaptQ, and RP-AdaptQ. We also consider a benchmark allocator that exploits rate predictions as in \ac{PGS}. However, it is energy independent, and its objective is to maximize user quality. This is achieved by solving \eref{eq:PGS-MinAir} with the objective of maximizing $q_{i,s}$. This allocator serves as reference to what can be achieved with rate predictions, but without considering energy savings, and is referred to as MaxQuality-MILP.

The network-wide video quality and power consumption metrics are defined as:
\begin{itemize}
\item $\tn{Q}_\tn{Net}$: the total quality of all delivered segments, divided by the number of requested segments.  
\item $\tn{F}_\tn{Net}$: the average percentage of playback time where the video is stalled, over all users.
\item $\tn{P}_\tn{Net}$: the average downlink power consumption of all \acp{BS} over the time window $T$.
\end{itemize}

\begin{figure}[!t]
	\noindent
	\centering
	\includegraphics[trim=0.0cm 0cm 0cm 0cm, clip=true, width=85mm]{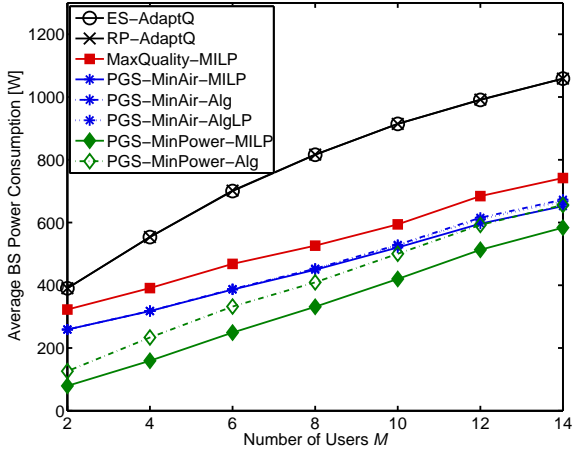} 
	\caption{BS downlink power consumption for varying number of users in the single-cell scenario.}
	\label{fig:PvsN_SC}
\end{figure}

\subsection{Single-cell Scenario}
\fref{fig:PvsN_SC} shows the average BS power consumption versus the number of users $M$. As expected, the allocators consume more power with increasing $M$. The PGS-MinAir schemes achieve significant power savings by exploiting rate predictions, without having to power down the BSs. The energy gains can also be viewed as spectral efficiency gains, where the saved airtime can be used for other users or applications. The MinAir-MILP and MinAir-AlgLP exhibit very close performance. This demonstrates the effectiveness of the developed multi-stage \ac{PGS} framework that is able to achieve close to optimal performance with significant complexity reduction. Also note that the MinAir-AlgLP and the MinAir-Alg achieve very close performance, and therefore the \ac{LP} formulation of \eref{eq:PGS-MinAirSeg} can be replaced with the simple segment airtime minimization algorithm presented in \sref{sec:AirMin}, without performance loss. The MinPower-MILP scheme saves additional power by switching the BS to sleep intermittently and making bulk transmissions to users when awake. The sleep times are coordinated such that the users' \ac{QoS} is not violated. When few users are present, the BS can sleep for prolonged periods of time and therefore the power savings can be very large (approximately one eighth of the baseline allocator power is needed). However, as expected, for larger $M$, MinPower-MILP gradually converges to MinAir-MILP since with many users the BS cannot find sufficient time for a `sleep session'. The MinPower-Alg performs close to the MinPower-MILP (exact solution) at low $M$, but then deviates and converges to MinAir-Alg. The reason is that MinPower-MILP jointly optimizes BS on/off states with BS airtime minimization, and is therefore able to strike the optimal tradeoff between serving users when their individual rates are high, and grouping user transmissions together (even if not at their respective best rates), to generate blocks of sleep time. This, however, comes at the cost of a tightly coupled \ac{MILP} that can take several minutes to solve. 

\begin{figure}[!t]
	\noindent
	\centering
	\includegraphics[trim=0.0cm 0cm 0cm 0cm, clip=true, width=85mm]{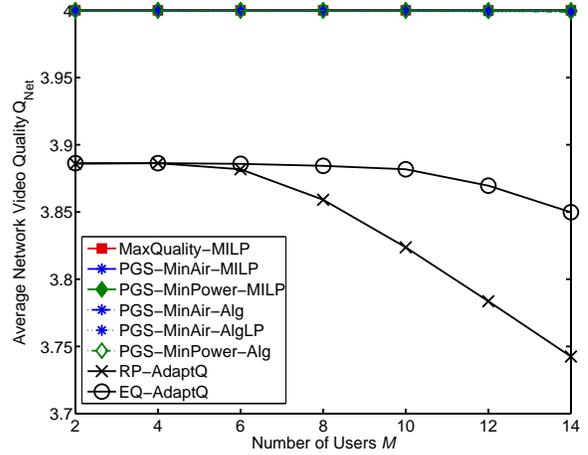} 
	\caption{Average quality level for varying number of users in the single-cell scenario.}
	\label{fig:QvsN_SC}
\end{figure}

In \fref{fig:QvsN_SC} we show the corresponding average segment quality level in this scenario. While the rate predictive schemes all achieve the highest quality of $4$, the baseline schemes experience a slight quality degradation as $M$ increases, with the RP-AdaptQ scheme suffering more. The video freezing, which is not depicted, was less than $1\%$ for all allocators.

\begin{figure}[!t]
	\noindent
	\centering
	\includegraphics[trim=0.0cm 0cm 0cm 0cm, clip=true, width=85mm]{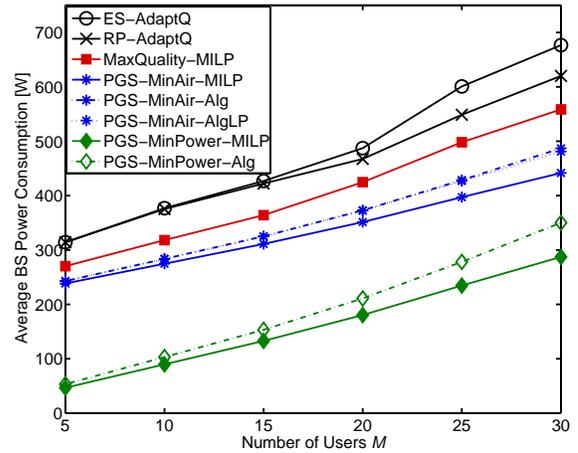} 
	\caption{BS downlink power consumption for varying number of users in the multi-cell highway scenario.}
	\label{fig:PvsN_Plat}
\end{figure}
\begin{figure}[!t]
	\noindent
	\centering
	\includegraphics[trim=0.0cm 0cm 0cm 0cm, clip=true, width=85mm]{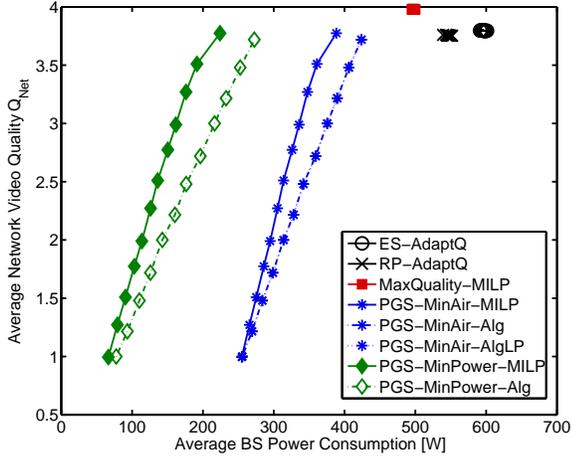} 
	\caption{The trade-off between the average video quality and the BS power consumption in the multi-cell highway scenario.}
	\label{fig:QvsP_Plat}
\end{figure}
\begin{figure}[!b]
	\noindent
	\centering
	\includegraphics[trim=0.0cm 0cm 0cm 0cm, clip=true, width=85mm]{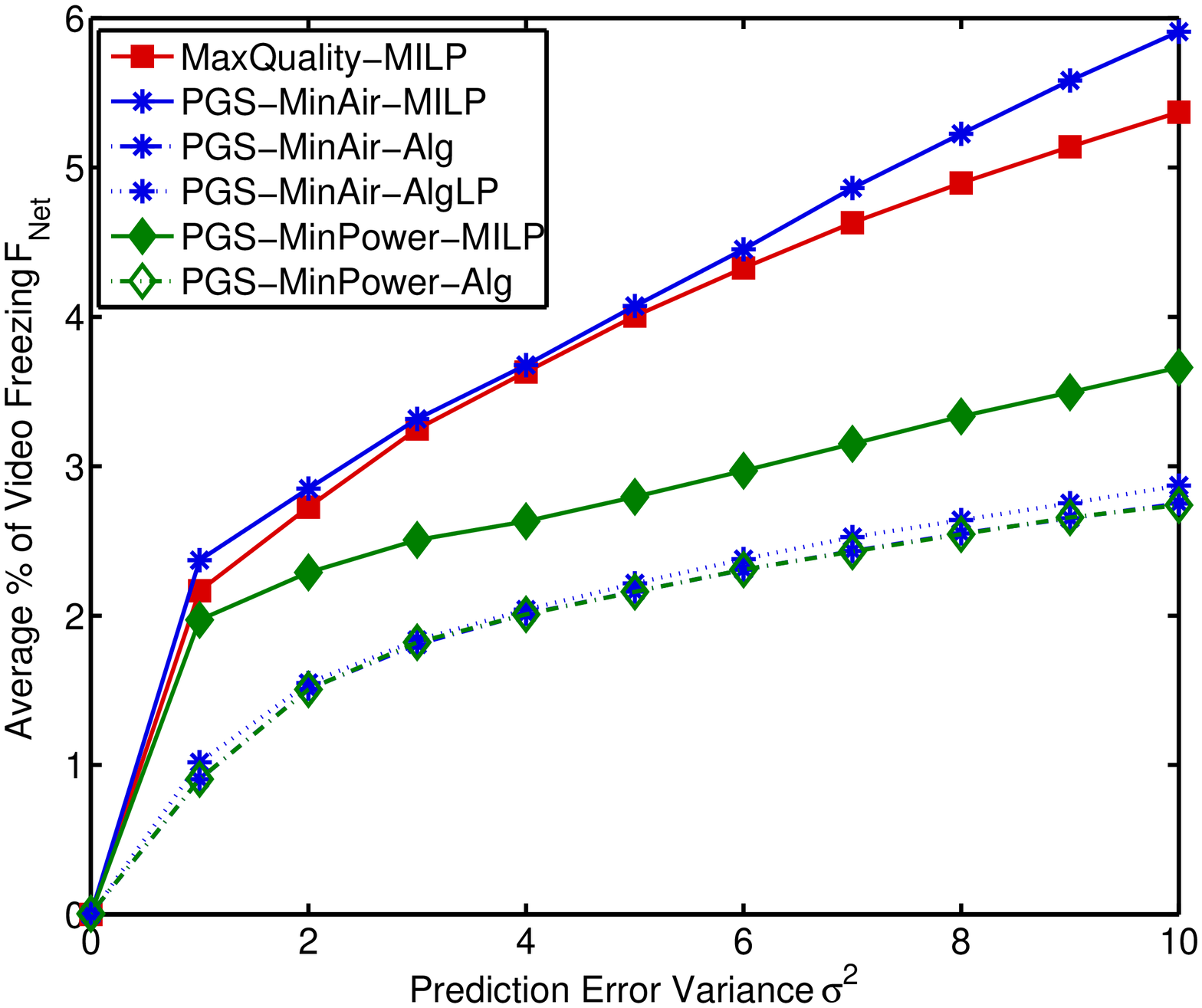} 
	\caption{Effect of prediction errors on video freezing of the  \ac{PGS} schemes in the highway scenario, $M=20$.}
	\label{fig:FvsVar_Plat}
\end{figure}
\begin{figure}[!b]
	\noindent
	\centering
	\includegraphics[trim=0.0cm 0cm 0cm 0cm, clip=true, width=85mm]{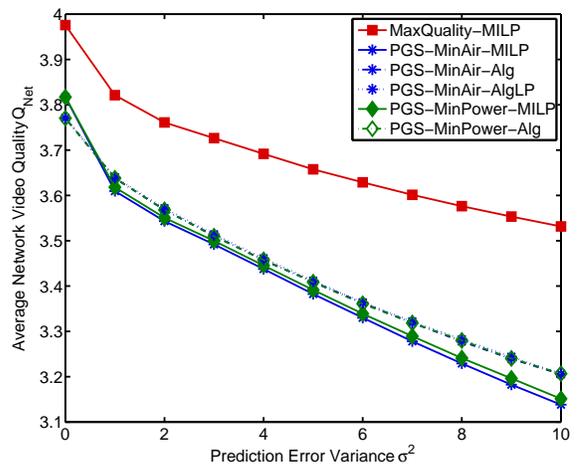} 
	\caption{Effect of prediction errors on the average video quality of the \ac{PGS} schemes in the highway scenario, $M=20$.}
	\label{fig:QvsVar_Plat}
\end{figure}

\subsection{Multi-cell Highway Scenario}
\fref{fig:PvsN_Plat} shows average BS power consumption versus the number of users $M$ for the three BS network of \fref{fig:Highway}. In this multi-cell scenario, the power saving potential of the MinPower-MILP scheme is observed, while all the allocators achieve an average quality level of 3.75. 
User mobility information allows the BSs to sleep before users arrive in the cells. Further, as the allocation plan is made over three cells, a user may be granted all the video content in one or two of the BSs and nothing in the third (i.e. allowing it to sleep). From \fref{fig:PvsN_Plat} we also note that the MinAir-Alg approaches deviate slightly from the MinAir-MILP solutions as $M$ increases. With many users in a multi-cell network, the problem is more complex and achieving optimality with the two-step rate allocation algorithm is more difficult. A similar observation can be made for MinPower-Alg. 

\fref{fig:QvsP_Plat} shows the tradeoff that the \ac{PGS} framework offers for video quality versus average BS power consumption. As illustrated, the power consumption of MinPower-MILP can be reduced by over $50\%$ as the quality is decreased. The MinAir-MILP scheme also offers significant power reduction, albeit at a lower ratio. Note that the deviation of the multi-stage algorithm solutions from the MILP solutions increases with quality, where again a higher load makes its more difficult to achieve optimality.

\subsection{Effect of Rate Prediction Errors}
To evaluate the effect of rate prediction errors on the \ac{PGS} schemes, we add a Gaussian random variable with a mean of zero and a variance $\sigma^2$ to the predicted user \ac{SNR}. 
The resulting user rate matrix is denoted by $\mathbf{\tilde r}$. Therefore, while the \ac{PGS} schemes use $\mathbf{\hat r}$ to minimize power, the actual rates received are determined by $\mathbf{x}\odot\mathbf{\tilde{r}}$. This can degrade video quality and cause video freezing if the resulting allocation does not completely download the segments in their due time.
\fref{fig:FvsVar_Plat} illustrates the impact of such errors on the video freezing for an increasing error variance $\sigma^2$. As expected, a higher error variance increases the video stalls. Interestingly, the algorithm-based \ac{PGS} schemes are more robust to prediction errors, and achieve under $3\%$ freezing for a high error variance. This indicates that even trends in the future user rates can provide significant power gains with minimal \ac{QoS} loss. There are two main reasons behind the larger \ac{MILP} solution sensitivity to prediction errors. First, since the \ac{PGS}-MILPs provide the lowest total airtime, when the observed rates are less than predicted, the user will be allocated an even lower rate, resulting in more freezing. This also explains why MinAir-MILP suffers more than MinPower-MILP (which has a larger airtime, but lower power due to the sleep modes). Secondly, the optimization based approaches make discrete allocation bursts, as observed in \fref{fig:BSPowerMinAir}. While being optimal, these bursts can be spaced out in time (to wait for user channel peaks). Therefore, when the actual rate is less than predicted, the user has to wait until the next allocation to resume playback. In contrast, the \ac{PGS} rate allocation algorithm performs allocation every $N_\tn{seg}$ slots, when a user does not have any buffered segments.

\begin{figure}[!t]
	\noindent
	\centering
	\includegraphics[trim=0.0cm 0cm 0cm 0cm, clip=true, width=85mm]{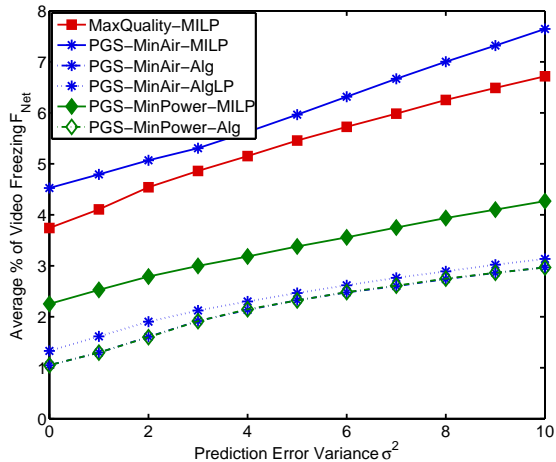} 
	\caption{Effect of prediction errors and fast fading on video freezing of the  \ac{PGS} schemes in the highway scenario, $M=20$.}
	\label{fig:FvsVarFF_Plat}
\end{figure}

\begin{figure}[!t]
	\noindent
	\centering
	\includegraphics[trim=0.0cm 0cm 0cm 0cm, clip=true, width=85mm]{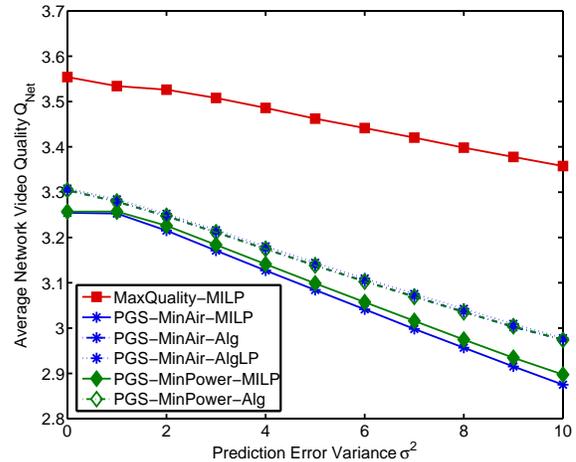} 
	\caption{Effect of prediction errors and fast fading on the average video quality of the \ac{PGS} schemes in the highway scenario, $M=20$.}
	\label{fig:QvsVarFF_Plat}
\end{figure}

In order to investigate the effect of fast fading, we model the channel with i.i.d. Rayleigh-fading as well. The resultant $\mathbf{\tilde r}$ is now computed from an SNR that has a Gaussian error component, and a fast-fading component.  The results are shown in \fref{fig:FvsVarFF_Plat} and \fref{fig:QvsVarFF_Plat}, where we can see that even with an error variance of zero, fast fading causes performance losses. Note that the relative effects of errors on the different PGS solutions follow similar trends to the previous results in \fref{fig:FvsVar_Plat} and \fref{fig:QvsVar_Plat}. 
To improve the performance under effects of fast fading we suggest that a more conservative measure of $\mathbf{\hat r}$ can be used while solving PGS. In other words, the values of $\mathbf{\hat r}$ can be decreased by a small factor to reduce the error effects on freezing, which happens when the actual rate is less than the predicted rate. 

\section{Conclusion}
\label{sec:Conc}
In this paper, we investigated how knowledge of future wireless data rates can improve spectral efficiency and provide downlink BS power savings. We used predicted rates to jointly optimize multi-user rate allocation, video segment quality, and BS on/off status. This was accomplished in an \ac{MILP} formulation that captures the user video streaming requirements, the \ac{BS} power consumption, and deep sleep mode operation. As the resulting \ac{MILP} can be computationally intractable for large problem sizes, a multi-stage polynomial-time algorithm was developed. Simulations demonstrate that high energy efficiency gains are achieved by the proposed adaptive video transmission framework. 
Our numerical results indicate that the proposed \ac{PGS} algorithms achieve close to optimal performance, while exhibiting a higher degree of \ac{QoS} robustness to prediction errors. 
Future work includes evaluating the potential of \ac{PGS} to prolong the battery life of \acp{UE} as well. The fact that our \ac{PGS} algorithm is less sensitive to prediction errors than the MILP formulation shows that there is room for further work. 
We, thus, plan to use stochastic channel models along with robust optimization techniques to improve the performance of PGS under uncertainty. 

\bibliography{IEEEabrv,refThesis-v2}

\begin{IEEEbiography}[{\includegraphics[width=1in,height=1.25in,clip,keepaspectratio]{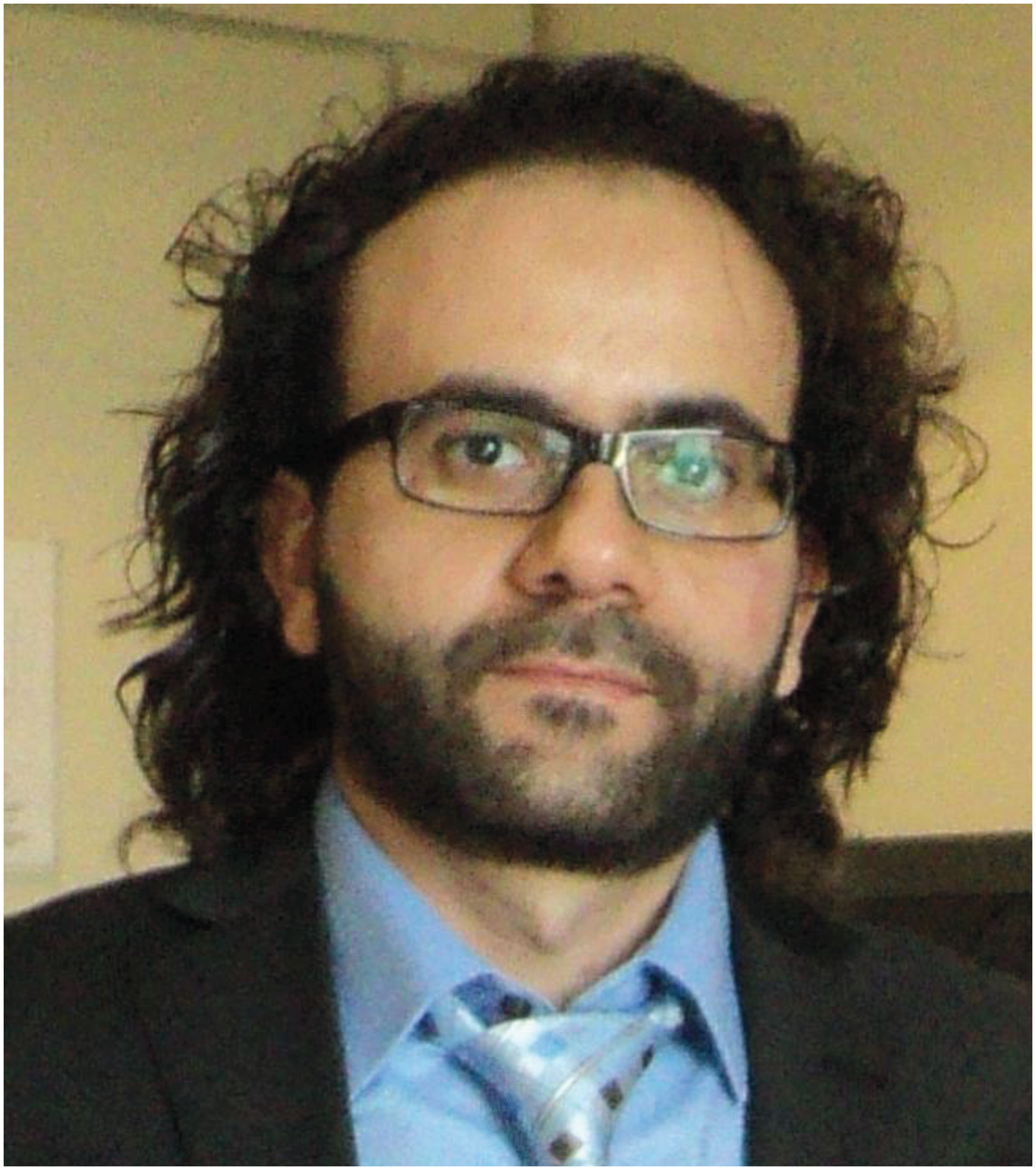}}]{Hatem Abou-zeid}
is a research assistant and PhD candidate in electrical and computer engineering at Queen's University, Canada.
He received his B.Sc. and M.Sc. degrees (with honors) in communications engineering from the Arab Academy for Science, Technology and Maritime Transport, Egypt, in 2005 and 2008, respectively. 
He was awarded a DAAD RISE-Pro research scholarship in 2011 for a six month internship at Bell Labs, Germany. 
His research interests include context-aware radio access networks, network adaptation and cross-layer optimization, adaptive video delivery, and vehicular communications.
Hatem is an experienced lecturer and has been granted several Teaching Fellowships at Queen's University to instruct freshman and senior-level engineering courses. 
He is also a technical reviewer for several prestigious conferences and journals.

\end{IEEEbiography}
\begin{IEEEbiography}[{\includegraphics[width=1in,height=1.25in,clip,keepaspectratio]{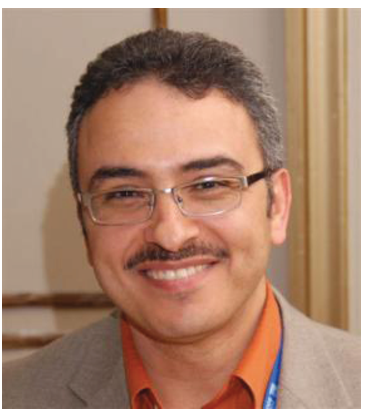}}]{Hossam Hassanein}
is a leading authority in the areas of broadband, wireless and mobile networks architecture, protocols, control and performance evaluation. His record spans more than 500 publications in journals, conferences and book chapters, in addition to numerous keynotes and plenary talks in flagship venues. Dr. Hassanein has received several recognition and best papers awards at top international conferences. He is also the founder and director of the Telecommunications Research (TR) Lab at Queen's University School of Computing, with extensive international academic and industrial collaborations. He is a senior member of the IEEE, and is a past chair of the IEEE Communication Society Technical Committee on Ad hoc and Sensor Networks (TC AHSN). Dr. Hassanein is an IEEE Communications Society Distinguished Speaker (Distinguished Lecturer 2008-2010).
\end{IEEEbiography}
\begin{IEEEbiography}[{\includegraphics[width=1in,height=1.25in,clip,keepaspectratio]{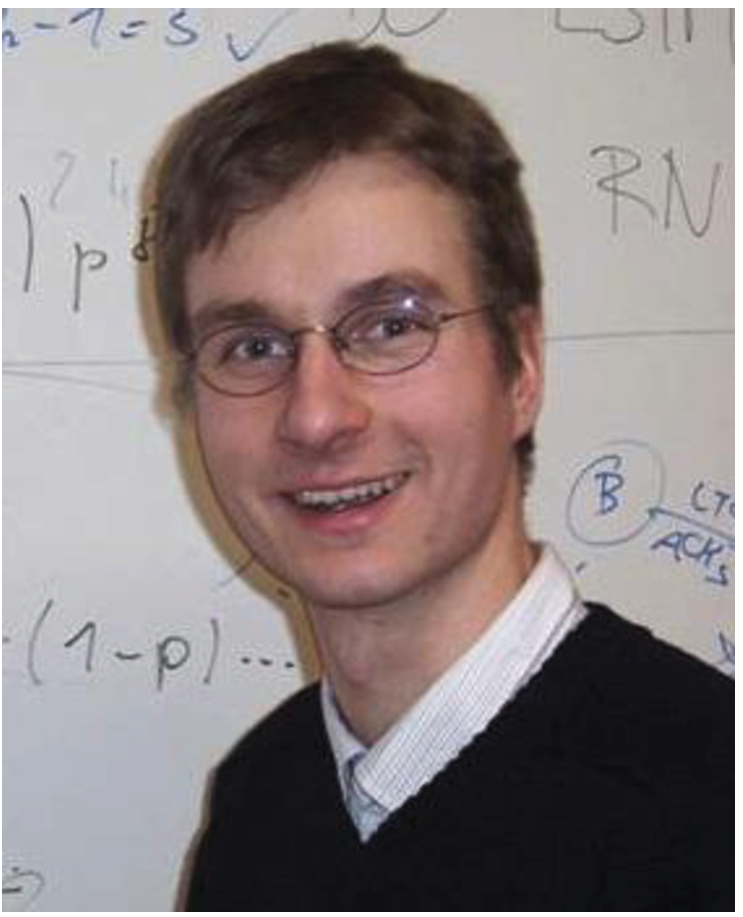}}]{Stefan Valentin}
has been a full researcher at Bell Labs, Stuttgart, Germany since 2010. Previous appointments include the University of Paderborn, Germany, the International Centre of Theoretical Physics, Italy, and the Carleton University, Canada. Stefan's main research interest is wireless resource allocation with Context-Aware and Anticipatory Adaptation in particular. In these fields, he leads several research projects and an integration activity with a market-leading operator. 
Stefan received a summa cum laude doctorate in Computer Science from the University of Paderborn in 2010, the SIMUTools Best Paper Award in 2008, the Klaus Tschira Award for Comprehensible Science in 2011, and the Bell Labs Special Award of Excellence in 2013. He is a member of the Alcatel-Lucent Leadership program since 2012, advises the German Federal Ministry of Education and Research (BMBF), and leads the PhD Internship program at Bell Labs Germany.
\end{IEEEbiography}

\end{document}